\documentclass[conference,compsoc]{IEEEtran}
\IEEEoverridecommandlockouts
\usepackage{hyperref}       
\usepackage{url}            
\usepackage{amsfonts}       
\usepackage{graphicx}       
\usepackage{amsthm}
\usepackage{amsmath}
\usepackage{algpseudocode, algorithm}
\usepackage{subcaption}
\algnewcommand\INPUT{\item[\textbf{Input:}]}%
\algnewcommand\OUTPUT{\item[\textbf{Output:}]}%
\DeclareMathOperator*{\argmax}{arg max}
\DeclareMathOperator*{\argmin}{arg min}

\ifCLASSOPTIONcompsoc
  \usepackage[nocompress]{cite}
\else
  \usepackage{cite}
\fi
\ifCLASSINFOpdf
\else
\fi
\begin{document}
\newtheorem{theorem}{Theorem}[section]
\newtheorem{lemma}{Lemma}[section]
\newtheorem{definition}{Definiton}[section]
\newtheorem{corollary}{Corollary}[section]
\newtheorem{proposition}{Proposition}[section]
\newtheorem{conjecture}{Conjecture}[section]
\newtheorem{example}{Example}

%
\title{Actual Knowledge Gain as Privacy Loss in Local Privacy Accounting}

\author{\IEEEauthorblockN{Mingen Pan}
\IEEEauthorblockA{Independent Researcher \IEEEauthorrefmark{1} \thanks{\IEEEauthorrefmark{1} Currently working at Google. This work was conducted independently and does not reflect the views or endorsement of Google.}\\
Email: mepan94@gmail.com}
}


%


\maketitle

\begin{abstract}

This paper establishes the equivalence between Local Differential Privacy (LDP) and a global limit on learning any knowledge specific to a queried object. However, an output from an LDP query is not necessarily required to provide exact amount of knowledge equal to the upper bound of the learning limit. The LDP guarantee can overestimate the amount of knowledge gained by an analyst from some outputs. To address this issue, the least upper bound on the actual knowledge gain is derived and referred to as \textit{realized privacy loss}. This measure is also shown to serve as an upper bound for the actual $g$-leakage in quantitative information flow.

The gap between the LDP guarantee and realized privacy loss motivates the exploration of a more efficient privacy accounting for fully adaptive composition, where an adversary adaptively selects queries based on prior results. The Bayesian Privacy Filter is introduced to continuously accept queries until the realized privacy loss of the composed queries equals the LDP guarantee of the composition, enabling the full utilization of the privacy budget of an object. The realized privacy loss also functions as a privacy odometer for the composed queries, allowing the remaining privacy budget to accurately represent the capacity to accept new queries. Additionally, a branch-and-bound method is devised to compute the realized privacy loss when querying against continuous values. Experimental results indicate that Bayesian Privacy Filter outperforms the basic composition by a factor of one to four when composing linear and logistic regressions.

\end{abstract}


%
\IEEEpeerreviewmaketitle

\section{Introduction}

Artificial intelligence has experienced a recent surge in development, largely driven by training models using vast amounts of data collected everywhere, including from individuals. Naturally, this raises concerns about privacy when personal data are utilized in AI models \cite{shokri2015privacy}. To address these concerns, Differential Privacy (DP), firstly proposed by \cite{dwork2006differential}, is proved to be able to protect individual privacy during model training \cite{shokri2015privacy, abadi2016deep}. DP ensures that, when a database is queried, the participation of an entry in the database point does not significantly alter the probability of any output. The original form of DP, also known as Centralized Differential Privacy (CDP), operates under the assumption that a trusted curator is responsible for aggregating data and ensuring that the output adheres to the DP guarantee. However, centralized data processing presents two significant problems: (1) users have to trust a curator, and (2) there are various potential vulnerabilities that may compromise user data, from eavesdropping on data transmission to hacking the curator's database \cite{hassan2019differential}.

To address these concerns, Local Differential Privacy (LDP) has been introduced by \cite{LDP}, whereby any data inputted into a query yields a similar output distribution. LDP is often achieved by directly perturbing the input. LDP offers the advantage of conducting queries locally on an individual's device, with only the perturbed result being transmitted. Theoretically, no third parties are able to acquire the original value of a user's data. As a result, tech giants, including Google \cite{rappor}, Apple \cite{apple_privacy}, and Microsoft \cite{ding2017collecting}, have embraced LDP as an integral part of their data processing protocols.

\subsection{Motivation}

Suppose an adversary has a prior belief about an object, derived either from the prior distribution of the object or from a subjective\footnote{One is allowed to have any prior belief as long as it is consistent with known observations and knowledge (Section \ref{sec:belief}). With enough effective observations, any prior belief will eventually converge to the true value \cite{bernardo2009bayesian}.} belief, and performs an LDP query against an object. Following Bayes' rule, the adversary's posterior belief about the object's value will remain similar to their prior belief. However, the definition of LDP does not explicitly guarantee the extent to which an adversary could learn or infer about an object. For instance, an LDP query may inquire about a person's age, and while it is improbable for an adversary to significantly update their belief about the age, can we also ensure that the adversary is unlikely to learn anything about whether the person has a certain disease? To address this gap, this paper introduces a learning limit on gaining knowledge about statements. Informally, a learning limit prevents the posterior belief $Q(f | y)$ of an adversary about a statement $f$ from changing too much compared to the prior belief $Q(f)$ after a query returns $y$. Subsequently, Section \ref{sec:equivalance} demonstrate that LDP is equivalent to providing a global learning limit for all statements specific to the queried object, and vice versa, formulated as:

\begin{equation*}
  \forall Q, f, y :   | \log \frac{Q(f | y)}{Q(f)} | \le \epsilon ,
\end{equation*}

\noindent where $y$ is an $\epsilon$-LDP query. This equivalence can be extended to all statements specific to the queried object and other objects conditionally independent of the query given the queried object.

However, while LDP provides a learning limit, it does not necessarily mean that an adversary can always gain knowledge exactly as the limit from every output of an LDP query. Consider the following example:

\begin{example}
    Suppose we have a binary object $X$ with value $x \in \{0, 1\}$, and a randomized response $M_{RR}$  that outputs $y \in {0, 1}$ with the following probability distribution:

    \begin{equation}
        Pr(y | x) = 
        \begin{cases}
            p, \text{ if } y = x, \\
            1 - p, \text{ otherwise}.
        \end{cases} \notag
    \end{equation}

     \noindent where $p > \frac{1}{2}$. An adversary executes $M_{RR}$ on $X$ for $k$ iterations, which happens to output the same number of ones and zeros, e.g., $[0, 1, 1, 0]$.
     \label{ex_1}
\end{example}

\noindent If the adversary applies Bayes' rule, their belief about the value of the object $X$ remains unchanged. Section \ref{sec:privacy_loss} demonstrates that under these conditions, a Bayesian adversary will not become more confident about any statement specific to this object given this output. On the other hand, $M_{RR}$ satisfies $\log \frac{p}{1-p}$-LDP \cite{wang2016using}. This paper mainly focuses on $\epsilon$-LDP, and the basic composition yields $k \log \frac{p}{1-p}$-LDP for the described process\cite{dwork2014algorithmic}, which is proven to be a tight bound for $\epsilon$-LDP\footnote{Note that $\frac{Pr(11...1 | X = 1)}{Pr(11...1 | X = 0)} = (\frac{p}{1-p})^k$. According to the definition of LDP (Section \ref{sec:LDP}), $\epsilon$ is at least $k \log \frac{p}{1-p}$.}. Thus, an adversary may still learn nothing from the queries with non-zero LDP guarantee.  The LDP guarantee $\epsilon$ is considered as the \textit{worst-case} privacy loss by literature like \cite{dwork2014algorithmic}, purely based on its definition $\max_{x, x', y} \frac{Pr(y|x)}{Pr(y|x')}$. However, it lacks the granularity to represent the actual privacy loss of a specific output and does not incorporate semantics related to knowledge gain. This discrepancy motivates the development of a more fine-grained and interpretable metric of privacy loss that connects actual knowledge gain and the DP guarantee.

Given that the actual knowledge gain depends on the output of a query, this paper proposes a privacy loss metric $L(X|y)$ mapping an output $y$ to the maximum actual knowledge gain, specifically the maximum change in an adversary's confidence about any statement specific to an object $X$ given $y$. The result of $L(X|y)$ given an output $y$ is referred to as \textit{realized} privacy loss. Subsequently, the DP guarantee is shown to impose an upper bound on the realized privacy loss of any output, i.e., $\max_y L(X|y) \le e^{\epsilon}$, consistent with its existing interpretation as the worst-case privacy loss. With this new metric for privacy loss, the author further defines privacy-budget utilization as the ability of a query to realize the privacy loss of an object. A query fully utilizes its privacy budget if and only if every output of the query could realize the privacy loss exactly as the DP guarantee. The composition of queries is equivalent to a complex DP query, so the privacy-budget utilization is also applicable to the composition.  In Example \ref{ex_1}, the basic composition does not fully utilize its privacy budget, prompting this paper to seek a more efficient composition that maximizes privacy-budget utilization.

\subsection{Result}
This paper focuses on the most flexible form of interaction between an analyst (adversary) and an object, where an analyst adaptively determines the next query based on the results of previous queries, referred to as fully adaptive composition. While many LDP algorithms (e.g., \cite{wang2016using, nguyen2016collecting}) query each object in a group only once, an object may receive repeated queries for different types of information throughout their lifetime, which is equivalent to an adaptive composition. For example, individuals' healthcare data may be queried repeatedly to analyze various diseases, or images may be queried continuously to extract new features. Moreover, training a machine learning model often requires querying the same object multiple times \cite{shokri2015privacy}. Notably, the LDP guarantee of the ``lifetime" composition should equal the privacy budget assigned to the object.

In this paper, privacy accounting refers to any algorithm designed to quantify the privacy loss of a composition and to determine whether the composition violates DP guarantee. \cite{rogers2016privacy} proposed a privacy accounting framework for fully adaptive composition, consisting of two components: (1) a privacy filter deciding whether to execute a query in order to satisfy the DP guarantee, and (2) a privacy odometer that monitors the privacy loss of the executed queries. This study adopts and extends these concepts to the context of LDP. Section \ref{sec:privacy_filter} demonstrates that the $\epsilon_g$-LDP guarantee for a composition is satisfied if the realized privacy loss does not exceed $e^{\epsilon_g}$. To this end, a Bayesian Privacy Filter (Algorithm \ref{algo:bayesian_privacy_filter}) is implemented to continuously accept new queries if every potential output from the next query still guarantees that the realized privacy loss will not exceed $e^{\epsilon_g}$. A simplified version of the Bayesian Privacy Filter is also introduced as Algorithm \ref{algo:simplified_bayesian_filter}, sharing a similar methodology but with reduced computational complexity. Furthermore, Section \ref{sec:aldp} extends the Bayesian Privacy Filter to $(\epsilon, \delta)$-LDP. Additionally, using the realized privacy loss as a privacy odometer is analyzed in Section \ref{sec:odometer}, allowing the remaining privacy budget to accurately represent the capacity to accept new queries. Finally, Section \ref{sec:role_of_bayesian_belief} discusses the role of Bayesian belief in interpreting the Bayesian Privacy Filter. Although Bayesian belief is employed for interpretation, the correctness of the privacy filter does not depend on the adoption of Bayesian belief.

The Bayesian Privacy Filter and Odometer require the realized privacy loss of the executed queries, whose computation can be challenging for an object with a domain of continuous values. Section \ref{sec:algo_to_compose} suggests a novel approach by translating the computation of realized privacy loss into a form of Difference of Convex (DC) Programming \cite{tao1997convex}. It further outlines the translating process for linear regression, truncated linear regression, and logistic regression. Additionally, a dedicated branch-and-bound method is introduced specifically for computing the realized privacy loss associated with the aforementioned regressions.

Eventually, Section \ref{sec:evaluation_of_bayesian_composition} empirically evaluates the composition managed by a Bayesian Privacy Filter, referred to as Bayesian composition, demonstrating that Bayesian composition is five times and two times as efficient as the basic composition for linear and logistic regressions, respectively. Real-world query composition is also evaluated with four regressions against healthcare data. With a 50\% chance, these queries consume less than 61\% of the total privacy budget when using the Bayesian composition.

\subsection{Related Work}
\label{sec:related_work}

Previous research \cite{lee2012differential, li2013membership} has examined privacy within the framework of Bayesian belief, identifying an adversary's belief regarding the identity and existence of an object as a significant privacy concern. In addition to DP, quantitative information flow (QIF)\cite{QIF} is another prominent framework to analyze privacy loss, referred to in QIF as information leakage. QIF measures the leakage as the reduction in entropy of an adversary's belief about an object's identity. Previous studies (e.g., \cite{barthe2011information, alvim2015information, chatzikokolakis2019comparing}) have indicated that DP imposes an upper bound on the max-case information leakage. Traditional information leakage primarily focuses on the identity of an object. However, ISO has published a standard for information security, which defines personally identifiable information (i.e., privacy of a person) as ``any information that is directly or indirectly linked to a person" \cite{ISO27001}. To address this, \cite{g-QIF} introduced $g$-leakage, which generalizes entropy reduction to account for any information about an object. Section \ref{sec:relationship_with_QIF} establishes that the realized privacy loss in Eq. \eqref{eq:privacy_loss} serves as an upper bound for all realized $g$-leakages. Consequently, the DP guarantee also bounds any max-case $g$-leakages and $g$-capacities, consistent with \cite{QIF_LDP}. Furthermore, Section \ref{sec:application_to_QIF} demonstrates that a Bayesian Privacy Odometer offers an upper bound for any realized $g$-leakage during an adaptive composition.

Meanwhile, previous research (e.g., \cite{dwork2016concentrated, rogers2016privacy, triastcyn2020bayesian}) has introduced the concept of a privacy loss random variable similar to the realized privacy loss. However, these studies have primarily focused on expected or worst-case privacy loss. Likewise, most research connecting DP and QIF (e.g., \cite{barthe2011information, alvim2015information, chatzikokolakis2019comparing, QIF_LDP}) has only analyzed expected or worst-case information leakage from DP queries. In contrast, this paper focuses on the realized privacy loss of executed queries, including its interpretation and application in adaptive composition.

Additionally, this paper has investigated the fully adaptive composition, which was initially proposed by \cite{rogers2016privacy}, and subsequently developed by \cite{feldman2021individual, lecuyer2021practical, whitehouse2022fully}. It has been applied in various scenarios \cite{abadi2016deep, durfee2019practical, lecuyer2019privacy}. However, all the existing works solely focused on CDP, and this paper stands as the first to explore the fully adaptive composition under the context of LDP.

\section{Background}

\subsection{Belief and Bayesian Inference}
\label{sec:belief}

Individuals hold unique beliefs regarding the actual world. Formally, a person's belief $Q:\mathcal{S} \rightarrow [0, 1]$ is defined as a function mapping a statement $s \in \mathcal{S}$ to the person's confidence about $s$.  Given a possible world $\omega$, $Q(W = \omega | \theta)$, abbreviated as  $Q(\omega | \theta)$, represents the confidence of a person believing that $\omega$ corresponds to the actual world $W$, where $\theta$ represents the auxiliary information possessed by the person. Although beliefs can be subjective \cite{bernardo2009bayesian}, $Q(\omega | \theta)$ should still satisfy the following properties:

\begin{enumerate}
    \item  $\forall \omega \in \Omega : Q(\omega | \theta) \ge 0$, and $\sum_{w \in \Omega} Q(\omega | \theta) = 1$, where $\Omega$ represents all possible worlds.
    \item If $\theta$ includes the marginal probability of an event $A$, denoted as $Pr(A)$,  
    \begin{equation}
        \sum_{\omega \in \Omega } \mathbf{1}\{ A \in \omega \} Q(\omega | \theta) = Pr(A) , \notag
    \end{equation}
    where $\mathbf{1}\{ A \in \omega \}$ returns one if the event $A$ occurs in the world $\omega$, otherwise zero.
    \item If $\theta$ includes the relationship between two events $A$ and $B$, represented as $Pr(B | A)$, then 
    \begin{equation}
         \frac{ \sum_{\omega \in \Omega } \mathbf{1}\{ B \in \omega \} \mathbf{1}\{ A \in \omega \} Q(\omega | \theta)}{ \sum_{\omega \in \Omega } \mathbf{1}\{ A \in \omega \} Q(\omega | \theta)} = Pr(B | A) . \notag 
    \end{equation}
\end{enumerate}

Subsequently, a person's confidence regarding an object $X$ being $x$ can be derived from the belief regarding all possible worlds $\{ Q(\omega | \theta) | \forall \omega \in \Omega \}$ as

\begin{equation}
        Q(X = x | \theta) = \sum_{\omega \in \Omega } \mathbf{1}\{ X = x | \omega \} Q(\omega | \theta) , \notag
\end{equation}

\noindent where $\mathbf{1}\{ X = x | \omega \}$ returns one if $X = x$ is true in the world $\omega$, otherwise zero. Subsequently, we have $\sum_{x \in \mathcal{X}} Q(X = x | \theta) = 1$, where $\mathcal{X}$ denotes all possible values of $X$. In this article, we will use the notation $Q(x | \theta)$ instead of $Q(X = x | \theta)$ when there is no ambiguity regarding the object we are referring to.

A Bayesian believer also updates their belief regarding an unknown object after making an observation. This update follows Bayes' theorem, which states:

\begin{equation}
    Pr(A|B) = \frac{Pr(B|A) Pr(A)}{Pr(B)},
    \label{eq:bayes}
\end{equation}

\noindent where $A$ and $B$ represent two events. The new belief (referred to as the posterior belief) of a Bayesian believer after observing an event $y$ is defined as:

\begin{equation}
    Q(x | \theta, y) = \frac{Pr(y | x) Q(x | \theta)}{\sum\limits_{x' \in \mathcal{X}} Pr(y | x') Q(x' | \theta)}.
    \label{eq:bayesian_belief}
\end{equation}

\noindent Eq. \eqref{eq:bayesian_belief} bears the same form as Eq. \eqref{eq:bayes} because $\sum\limits_{x' \in \mathcal{X}} Pr(y | x') Q(x' | \theta) = Q(y|\theta)$, which represents the person's belief regarding the marginal probability of the event $y$. In the subsequent article, we may omit $\theta$ in $Q(x | \theta, y)$ and $Q(x | \theta)$, thereby denoting them as $Q(x |y)$ and $Q(x)$, respectively, when the auxiliary information before an observation is not the main focus of interest.

\subsection{Probabilistic Knowledge}
\label{sec:prob_knowledge}

Probabilistic knowledge, initially proposed by \cite{bacchus1989representing}, has gained widespread acceptance as a means of representing uncertainty and confidence regarding statements (e.g., \cite{halpern2017reasoning}). It has found extensive applications in fields such as machine learning \cite{richardson2006markov, russell2015unifying}. To elucidate probabilistic knowledge, we follow the first-order logic \cite{barwise1977fol} to define a statement (proposition) specific to an object $X$ as applying a predicate $f$ to $X$ or its possible value $x$. For instance, $f(X)$ can represent a statement as ``$X$ is a die and will return one when rolled randomly". Given a possible value $x$, the evaluation of a traditional predicate will return true or false, while probabilistic knowledge allows the correctness of $f(x)$ to be a probability, denoted as $Pr(f(x))$. Since $f$ is a single-variable predicate, $Pr(f(x))$ depends only on the input $x$. Using the aforementioned example, if $x$ represents a perfect six-sided die, $Pr(f(x)) = \frac{1}{6}$. The confidence of a person regarding the truth of the statement $f(X)$ is defined as

\begin{equation}
    Q(f(X)) = \sum\limits_{x \in \mathcal{X}} Pr(f(x)) Q(X = x). \notag
\end{equation}

\noindent For instance, if a person believes that a die $X$ has a 50\% chance of being a perfect six-faced die and a 50\% chance of being a perfect four-faced die, their $Q(f(X))$ will be $0.5 \times \frac{1}{6} + 0.5 \times \frac{1}{4} = \frac{5}{24}$. We abbreviate $f(X)$ and $Q(f(X))$ as $f$ and $Q(f)$, respectively, when there is no ambiguity in referring to $X$. Analogous to the aforementioned posterior belief (Eq. \eqref{eq:bayesian_belief}), the confidence about a statement $f$ after an observation $y$ is given by
\begin{multline}
    Q(f | y) = \sum\limits_{x \in \mathcal{X}} Pr(f(x)) Q(x | y) \\ =  \frac{\sum\limits_{x \in \mathcal{X}} Pr(f(x)) Pr(y | x) Q(x)}{\sum\limits_{x' \in \mathcal{X}} Pr(y | x') Q(x')}. \notag
\end{multline}

When a person observes an event, the person may be more confident or more uncertain about a statement. Nevertheless, the person learns something new about the statement, so we call the change of confidence as knowledge gain about the statement, which is measured by $\frac{Q(f|y)}{Q(f)}$.

The discussion above can be extended to a statement specific to multiple objects by treating all the involved objects as a single ``composite" object.

\subsection{Quantitative Information Flow (QIF)}
\label{sec:QIF}
QIF\cite{QIF} utilizes information theory to analyze information leakage from a channel, referred to as a query in this paper. QIF interprets leakage as the increase in the vulnerability of a protected object. As noted in \cite{g-QIF}, vulnerability should not only account for the identity of an object but all information associated with the object. They propose a gain function $g(w \in \mathcal{W}, x \in \mathcal{X})$ to quantify the relevance between $x$ and $w$, where $\mathcal{W}$ represents a collection of information potentially related to the object $X$. Different gain functions correspond to different semantics. With a gain function $g$,  $g$-prior vulnerability of a protected object  $X$ is defined as

\begin{equation}
    V_g(Q) = \max_w \sum_{x \in \mathcal{X}} Q(X = x) g(w, x) ,
    \notag
\end{equation}

\noindent where $Q$ is the belief of a person. This paper is interested in the vulnerability given each output $y$ of a query against $X$. Thus, the realized $g$-posterior vulnerability is defined as

\begin{equation}
    V_g(Q | y) = \max_w \sum_{x \in \mathcal{X}} Q(X = x | y) g(w, x) .
    \notag
\end{equation}

\noindent Subsequently, the realized $g$-leakage is defined as 

\begin{equation}
    LK_g(Q | y) = \frac{V_g(Q | y)}{V_g(Q)} .
    \notag
\end{equation}

\subsection{Local Differential Privacy (LDP)}
\label{sec:LDP}

To protect the true value of an object from being inferred by an adversary, \cite{LDP} has proposed LDP to ensure a query to yield similar output distributions for all inputs.

\begin{definition}
   A query $M$ against an object $X$ is $\epsilon$-LDP, if and only if $\forall x, x' \in \mathcal{X},  y \in \text{Range}(M)$,
\begin{equation}
    Pr(M(X) = y | X = x) \le e^{\epsilon} Pr(M(X) = y | X = x') . \notag
\end{equation} 
\end{definition}

\noindent For convenience, $Pr(M(X) = y | X = x)$ may be abbreviated as $Pr(M(x) = y)$ or $Pr(y|x)$ if there is no ambiguity. The formula above is equivalent to:
\begin{multline}
\forall y \in \text{Range}(M): \\
    \max\limits_{x \in \mathcal{X}} Pr(M(x) = y) \le e^{\epsilon} \min\limits_{x \in \mathcal{X}}  Pr(M(x) = y) . \notag
\end{multline}

\noindent While there are other variations of LDP, this paper focuses only on the most common form mentioned above.

\subsection{LDP of Continuous Value}
\label{sec:LDP_of_continuous_value}

Consider a query against an object $X$ yielding an output $y \in [a, b]$. Depending on the purpose, various methods can be employed to enforce LDP for $y$. 

This paper specifically focuses on LDP enforcement for estimating the mean of $y$ from a group of objects. The chosen approach is a well-established perturbation algorithm, initially proposed by \cite{nguyen2016collecting} and later refined by \cite{ye2019privkv}. This algorithm discretizes and perturbs $y$, producing either $a$ or $b$ with the following probabilities:

\begin{equation}
    Pr(b | y) =  \frac{1}{b - a} \frac{ e^{\epsilon} - 1}{ e^{\epsilon} + 1} (y - a) +  \frac{1}{ e^{\epsilon} + 1} ,
    \label{eq:linear_pr}
\end{equation}

\noindent and $Pr(a | y) = 1 - Pr(b | y)$. Intuitively, if $y = a$, the probability to output $b$ is $\frac{1}{ e^{\epsilon} + 1}$. As $y$ approaches $b$, its probability to output $b$ increases linearly and eventually reaches $\frac{ e^{\epsilon}}{ e^{\epsilon} + 1}$.  Clearly, this algorithm adheres to $\epsilon$-LDP.

To estimate the mean of $y$ from a group of objects, the first step involves computing the frequency of $b$ outputted by the perturbation algorithm, denoted as $\pi_b$. Subsequently, the mean of $y$ can be estimated as:

\begin{equation}
    \bar{y} \approx a + b \frac{(e^\epsilon + 1) \pi_b - 1}{e^\epsilon - 1} . \notag
\end{equation}

\section{Learning from LDP queries}

LDP ensures that all objects have indistinguishable outputs. However, the definition of LDP does not explicitly restrict how much confidence an adversary can gain about any particular knowledge regarding an object. In this section, we will introduce a metric to quantify the amount of knowledge an adversary can acquire from a query and then demonstrate how LDP effectively prevents significant knowledge gain by the adversary.

\subsection{Learning Limit}
\label{sec:learning_limit}

Here, we will use $\frac{Q(f|y)}{Q(f)}$, the knowledge gain about a statement $f$ after an event $y$, to define a learning limit that imposes a constraint on the maximum amount of knowledge an adversary can acquire about an object.

\begin{definition}
    A query $M$ provides an $\epsilon$-learning limit for a statement $f$, if and only if, for any output $y$ from $M$ and any prior belief $Q$, the following inequality

    \begin{equation}
        e^{-\epsilon} \le \frac{Q(f | y)}{Q(f)} \le e^{\epsilon}
        \label{eq:learning_limit}
    \end{equation}
    \label{def:learning_limit}
    always holds true.

\end{definition}

When querying against an object, all statements specific to the object should be considered as privacy. For instance, in a survey, a participant may provide only basic biometric information such as age, height, and weight, without disclosing any medical history. However, a medical expert may still be able to infer with high confidence whether a person has a certain disease based on their biometric information. To prevent adversaries from gaining too much new knowledge about an object, it is crucial for privacy-preserving queries to maintain a reasonable learning limit for all statements specific to the object.

\subsection{Equivalence between LDP and Learning Limit}
\label{sec:equivalance}

This section aims to establish an equivalence between LDP and a global learning limit for all statements specific to a queried object:

\begin{lemma}
    For any statement $f$ specific to $X$, prior belief $Q$, and output $y$ from a query $M(X)$, the inequality
    \begin{equation}
        \frac{\min\limits_{x \in \mathcal{X}}Pr(M(x) = y)}{\max\limits_{x' \in \mathcal{X}}Pr(M(x') = y)} \le \frac{Q(f | y)}{Q(f)}  \le  \frac{\max\limits_{x \in \mathcal{X}}Pr(M(x) = y)}{\min\limits_{x' \in \mathcal{X}}Pr(M(x') = y)}
        \label{eq:bound_of_Q_f}
    \end{equation}
    holds true\footnote{Hereinafter, if $\min_{x' \in \mathcal{X}} Pr(M(x') = y) = 0$, the right-hand side of Eq. \eqref{eq:bound_of_Q_f} is assumed to be $+\infty$.}.
    \label{lem:bound_of_Q_f}
\end{lemma}

\begin{theorem}
    If a query $M$ is $\epsilon$-LDP, then $M$ also provides an $\epsilon$-learning limit for all statements specific to the queried object.
    \label{thm:ldp_to_learnig_limit}
\end{theorem}

\begin{lemma}
    For any output $y$ from a query $M$, we have
    \begin{equation}
        \sup_{Q, f} \frac{Q(f | y)}{Q(f)} = \frac{\max\limits_{x \in \mathcal{X}}Pr(M(x) = y)}{\min\limits_{x' \in \mathcal{X}}Pr(M(x') = y)}
        \label{eq:bound_of_Q_f_right_eq}
    \end{equation}
    holds true.
    \label{lem:Q_f_bound_eq}
\end{lemma}

\begin{theorem}
    If a query $M$ provides an $\epsilon$-learning limit for all statements specific to the queried object, then $M$ is $\epsilon$-LDP.
    \label{thm:learning_limit_to_LDP}
\end{theorem}

\noindent Their proofs are provided in Appendix \ref{sec:proof_LDP=learning_limit}.  Specifically, Theorems \ref{thm:ldp_to_learnig_limit} and \ref{thm:learning_limit_to_LDP} formally establish the equivalence between LDP and a global learning limit for all statements specific to a queried object. Additionally, Lemmas \ref{lem:bound_of_Q_f} and \ref{lem:Q_f_bound_eq} are presented here because they will also be used in the subsequent sections.

The equivalence can be extended to statements involving multiple objects, $\mathbf{X}$, including the queried object $X$, provided that $Pr(M(X)=y|\mathbf{X})$ still satisfies $\epsilon$-LDP for all possible values of $y$ and $\mathbf{X}$. This holds if $\mathbf{X}$ does not contain latent variables of $M$ or represent the future output of $M(X)$. The proof of this extension is analogous to those of the theorems above, with $X$ replaced by $\mathbf{X}$. 

\subsection{What is Privacy Loss?}
\label{sec:privacy_loss}

Theorem \ref{thm:ldp_to_learnig_limit} states that every output of an $\epsilon$-LDP query is allowed to change the confidence of an adversary about any statement specific to the queried object up to an $\epsilon$-learning limit. However, being allowed does not necessarily imply being able, and an LDP process sometimes may not even reveal privacy at all. Recall Example \ref{ex_1} above; we compute the likelihood function for both values:
\begin{multline}
    Pr(y_{\le k} = 0110 | X = 1) = Pr(y_{\le k} = 0110 | X = 0) \\ = p^{\frac{k}{2}} (1-p)^{\frac{k}{2}} . \notag
\end{multline}

\noindent Based on Lemma \ref{lem:bound_of_Q_f}, for any statement $f$ specific to $X$, $\frac{Q(f | y)}{Q(f)} = 1$. This implies that an adversary does not gain any knowledge about the object, and we must acknowledge that no privacy is compromised from the object. Therefore, instead of simply using the DP guarantee of a query to represent privacy loss, it is more appropriate to consider the measure of knowledge gain. Let us once again recall Lemma \ref{lem:bound_of_Q_f}. It states that for all statements specific to $X$, their knowledge gain can be bounded by a pair of symmetric bounds. Furthermore, Lemma \ref{lem:Q_f_bound_eq} demonstrates that these bounds are tight and can be approached by the knowledge gain of at least one statement. Thus, the least upper bound of knowledge gain can be utilized to quantify the privacy loss:

\begin{definition}
    Privacy loss $L(X | y)$ of an object $X$ after a query $M(X)$ returns $y$ is defined to be
    \begin{equation}
        L(X | y) = \frac{\max\limits_{x \in \mathcal{X}}Pr(M(x) = y )}{\min\limits_{x' \in \mathcal{X}}Pr(M(x') = y)} .
        \label{eq:privacy_loss}
    \end{equation}
    \label{def:privacy_loss}
\end{definition}

\noindent In the subsequent article, $L(X | y)$ may be abbreviated as $L(y)$ if there is no ambiguity about the queried object. $L(y)$ is also referred to as \textit{realized} privacy loss if the output $y$ is known. Let's recall the definition of LDP, which implies that $\forall y, L(y) \le e^\epsilon$. If $\epsilon$ is a tight bound for the query, it indicates $\exists y, L(y) = e^\epsilon$, representing the worst case of potential privacy loss (knowledge gain) from a query. Thus, the DP guarantee provided by a query can be seen as the \textit{worst-case} privacy loss. If we aim to limit the privacy loss for a future query, then the worst-case privacy loss is the metric to consider. However, when examining the privacy loss of a composition, we must consider both the realized privacy loss of the executed queries and the worst-case privacy loss of future queries.

Here, an example is presented to compute the realized privacy loss.

\begin{example}
\label{ex_2}
    (Running Example) An LDP-protected object $X$ has 11 possible values as 0, 1, 2, ..., 10. An analyst queries the values of $X$, $X^2$, and $X^3$. For any $i$, the LDP query against $X^i$ follows the output distribution:
    \begin{equation}
    \label{eq:runing_example_prob}
        Pr(y | x) = 
        \begin{cases}
            0.2 \times (\frac{x}{10})^i + 0.4, \text{ if } y = 1; \\
            1 - Pr(y = 1 | x), \text{ if } y = 0 .
        \end{cases} 
    \end{equation}
    One could verify that each query satisfies $\log(\frac{3}{2})$-LDP. Here, we illustrate how the realized privacy loss is computed when the queries return 1, 0, and 1. Notably, we have $Pr(y_1, y_2, y_3 | x) = \prod_{1 \le i \le 3} Pr(y_i|x)$, which is computed for every $x$ as
    \begin{center}
    \begin{tabular}{ | c | c |  }
    \hline
     x & $\prod_{1 \le i \le 3} Pr(y_i|x)$ \\ 
     \hline
     0 & 0.096   \\  
     1 & 0.101    \\  
     2 & 0.109   \\  
     ... &  \\
     10 & 0.144   \\  
     \hline
    \end{tabular}
    \end{center}

    The $\max$ and $\min$ of $\prod_{1 \le i \le 3} Pr(y_i|x)$ are 0.144 and 0.096, corresponding to $x = 10$ and $x = 0$, respectively. Thus, the realized privacy loss is 1.5.
\end{example}

With the realized privacy loss, we define the utilization of privacy budget as follows:

\begin{definition}
     An $\epsilon$-LDP query $M$ fully utilizes its privacy budget if and only if $L(X | y) = e^\epsilon$ holds for every output $y$ from $M$.
\end{definition}

\noindent That is, if a query fully utilizes its privacy budget, each of its outputs will realize the same privacy loss as the worst-case privacy loss. It is evident to see that Randomized Response \cite{wang2016using}, the fundamental algorithm in LDP, fully utilizes its privacy budget, though the proof is omitted here. In Example \ref{ex_1}, the basic composition does not fully utilize its privacy budget, as the realized privacy loss is zero. Example \ref{ex_2} also does not fully utilize its privacy budget , since the basic composition yields $3\log(\frac{3}{2})$-LDP ($e^\epsilon = 3.375$), while the realized privacy loss is just $1.5$. Based on these examples, this paper aims to find a more efficient composition algorithm that guarantees that the realized privacy loss could approach the DP guarantee as close as possible.

\subsection{Relationship with QIF}
\label{sec:relationship_with_QIF}

This section presents how the realized privacy loss and LDP are associated with QIF. A gain function $g(w, \cdot)$ in QIF can be translated to a statement $f_{g, w}(x)$ as ``A program receives a value $x$ and outputs one with a probability $g(w, x)$, otherwise zero. It returns one given an object $X$". Given $f_{g, w}(x) = g(w, x)$, we have the $g$-vulnerability as 
\begin{align*}
    V_g(Q) = \max_w Q(f_{g, w}) , \\ 
    V_g(Q | y) = \max_w Q(f_{g, w} | y) .
\end{align*}

\noindent  By combining these notions with Lemma \ref{lem:bound_of_Q_f}, we have:
\begin{theorem}
    \label{thm:realized_g_leakage_ub}
    The realized privacy loss $L(y)$ is an upper bound of any realized $g$-leakage, i.e., 
    \begin{equation*}
        \forall y: \max_{g, Q} LK_g(Q | y) \le L(y) .
    \end{equation*}
\end{theorem}

\begin{corollary}
    \label{cor:max_case_g_leakage_ub}
    If a query is $\epsilon$-LDP, we have
    \begin{equation}
        \max_{g, Q, y} LK_g(Q | y) \le e^{\epsilon} .
        \notag
    \end{equation}
\end{corollary}

\noindent The proofs are provided in Appendix \ref{sec:proof_qif}. Notably, Corollary \ref{cor:max_case_g_leakage_ub} implies that both the max-case $g$-leakage, i.e., $\max_y LK_g(Q | y)$,  and $g$-capacity, i.e., $\max_{Q,y} LK_g(Q | y)$, two widely used QIF metrics, are bounded by $e^{\epsilon}$. This result is consistent with\cite{QIF_LDP}.

This paper adopts $\frac{Q(f | y)}{Q(f)}$ to represent knowledge gain rather than $g$-leakage, as the semantics of $g$-leakage are undefined unless $g$ is specified. Additionally, a decrease in $\frac{Q(f | y)}{Q(f)}$ directly reflects an increase in confidence about the incorrectness of a statement. In contrast, the $\max$ operator in the definition of $g$-leakage (or vulnerability) complicates the interpretation of increased confidence in inverse cases.

\section{Privacy Accounting in Fully Adaptive Composition}

Privacy accounting encompasses two main functions \cite{rogers2016privacy}: (1) a privacy filter to ensure the DP guarantee of a composition and (2) a privacy odometer to monitor the privacy loss during the composition. In this section, we will focus on fully adaptive composition, where an adversary adaptively determines which query to execute based on the previous queries and their results. Fully adaptive composition is regarded as the most flexible form of adaptive composition, as it can represent any type of interaction between an adversary and an object, as long as the queries are executed sequentially. 

\subsection{Privacy Filter}
\label{sec:privacy_filter}

A privacy filter for LDP is a state machine that either accepts and rejects a query against an object, while ensuring the composition of all accepted queries does not exceed its DP guarantee. The interaction between an adversary and an object protected by a privacy filter appears as follows: the adversary collects all previous queries and their results, and is allowed to dry run the privacy filter to know if a query will be accepted. Considering all the information above, the adversary will either choose a valid query against the object or terminate the interaction. If a query is chosen, it will be verified by the privacy filter, executed against the object, and return a result. The adversary then repeats the aforementioned steps. Further details are presented in Algorithm \ref{algo:interaction}.

\begin{algorithm}
    \caption{Interaction between an adversary and an object (Fully Adaptive Composition)}
    \label{algo:interaction}
  \begin{algorithmic}[1]
    \INPUT Adversary $\mathcal{A}$, Privacy Filter $F$, and object $X$
    \State $i = 1$
    \Loop
        \State $\mathcal{A} = \mathcal{A}(F, X, M_1, y_1, M_2, y_2, ..., M_{i-1}, y_{i-1})$ \:\:  // $\mathcal{A}$ collects existing information
        
        \State $\mathcal{A}$ gives either $M_i$ or \textit{HALT}.  \:\: // $M_i$ will always be accepted by $F$ if given.
        \If{$\mathcal{A}$ gives \textit{HALT}}  \State\textbf{break}.
        \EndIf
        \State Assert $F$ accepts $M_i$. Otherwise, \textbf{continue}.
        \State $\mathcal{A}$ receives $y_i = M_i(x)$. \label{step:interaction_accept_query}
        \State $i = i + 1$
    \EndLoop
    \OUTPUT $(M_1, y_1, M_2, y_2, ..., M_{i-1}, y_{i-1})$
  \end{algorithmic}
\end{algorithm}

Now, we will formally define the DP guarantee of a privacy filter. The interaction $q(F, \mathcal{A}, X)$ between the adversary $\mathcal{A}$ and an object $X$ protected by a privacy filter $F$ can be regarded as a mechanism against an object (i.e., $M(\cdot) = q(F, \mathcal{A}, \cdot)$). Therefore, we can directly apply the definition of LDP to the interaction:

\begin{definition}
    An interaction $q(F, \mathcal{A}, \cdot)$ is $\epsilon$-LDP if and only if, for any $x, x' \in \mathcal{X}$, and any output $V = (M_1, y_1, ..., M_n, y_n)$ from the $q(F, \mathcal{A}, \cdot)$, the inequality 
    \begin{equation}
        Pr(q(F, \mathcal{A}, x) = V)  \le e^{\epsilon} Pr( q(F, \mathcal{A}, x') = V) 
        \label{eq:q_F_A_DP}
    \end{equation}
    holds.
\end{definition}

\noindent Consequently, the DP guarantee of a privacy filter can be defined as:

\begin{definition}
    A privacy filter $F$ is $\epsilon_g$-LDP\footnote{$\epsilon_g$ is used hereinafter to denote the global LDP guarantee (i.e., the privacy budget of an object), whereas $\epsilon$ is reserved for a single query.} if and only if $\forall \mathcal{A}$, $q(F, \mathcal{A}, \cdot)$ is  $\epsilon_g$-LDP.
\end{definition}

\noindent Based on Algorithm \ref{algo:interaction}, the choice of query $M_i$ only depends on the adversary $\mathcal{A}$, the privacy filter $F$, and previous queries $M_{<i} = \{M_j  | 1 \le j < i\}$ and their results $y_{<i} = \{y_j  | 1 \le j < i\}$. Therefore, its conditional probability is represented as $Pr(M_i | \mathcal{A}, F, M_{<i}, y_{<i})$. The result $y_i$ of query $M_i$ only depends on the queried object $X$. Suppose $X$ has a value of $x$. The probability of outputting $y_i$ is derived as $Pr(M_i(x) = y_i)$. Therefore, 
\begin{multline}
    Pr(q(F, \mathcal{A}, x) = V) = \\
    = \prod\limits_{1 \le i \le n} Pr(M_i | \mathcal{A}, F, M_{<i}, y_{<i}) Pr(M_i(x) = y_i) \\ \times Pr(\text{HALT} | \mathcal{A}, F, M_{\le n}, y_{\le n}) ,
    \label{eq:Pr_q_F_A_x}
\end{multline}

\noindent where $V = (M_{\le n}, y_{\le n})$ represents a valid output from $q(F, \mathcal{A}, x)$ and $n$ is the number of accepted queries in this output. Since $Pr(M_i | \mathcal{A}, F, M_{<i}, y_{<i})$ and $Pr(\text{HALT} | \mathcal{A}, F, M_{\le n}, y_{\le n})$ are independent of $x$, substituting Eq. \eqref{eq:Pr_q_F_A_x} into Eq. \eqref{eq:q_F_A_DP} yields
\begin{equation}
    \prod\limits_{1 \le i \le n} Pr(M_i(x) = y_i) \le e^{\epsilon_g} \prod\limits_{1 \le i \le n} Pr(M_i(x') = y_i) , \notag
\end{equation}

\noindent which is equivalent to

\begin{equation}
    \frac{\max\limits_{x \in \mathcal{X}} \prod\limits_{1 \le i \le n} Pr(M_i(x) = y_i)}{\min\limits_{x' \in \mathcal{X}} \prod\limits_{1 \le i \le n} Pr(M_i(x') = y_i)} \le e^{\epsilon_g}  .
    \label{eq:privacy_filter_dp}
\end{equation}

\noindent As a result, we just proved the following theorem:

\begin{theorem}
     A privacy filter $F$ is $\epsilon_g$-LDP if and only if, for any adversary $\mathcal{A}$ and any result $V = (M_1, y_1, ..., M_n, y_n)$ from the interaction $q(F, \mathcal{A}, \cdot)$, Eq. \eqref{eq:privacy_filter_dp} is satisfied.
     \label{thm:privacy_filter_ldp}
\end{theorem}

It is observed that the left-hand side of the above inequality is actually the realized privacy loss $L(y_{\le n})$, indicating that the realized privacy loss must not exceed the DP guarantee. A sufficient condition for a privacy filter to maintain $\epsilon_g$-LDP after accepting a query $M$ is that every output of $M$ still satisfies Eq. \eqref{eq:privacy_filter_dp}. Thus, we have the following corollary:

\begin{corollary}
    
    If, for any $y$ from a query $M$, 
    \begin{equation}
    \frac{\max\limits_{x \in \mathcal{X}} \prod\limits_{1 \le i \le n} Pr(M_i(x) = y_i) \times Pr(M(x) = y)}{\min\limits_{x' \in \mathcal{X}} \prod\limits_{1 \le i \le n} Pr(M_i(x') = y_i) \times Pr(M(x') = y)} \le e^{\epsilon_g}  
    \label{eq:privacy_filter_accept}
    \end{equation}
    \noindent holds true, where $M_{\le n}$ are the accepted queries, then a privacy filter accepting $M$ is still $\epsilon_g$-LDP. 
    \label{cor:privacy_filter_accept}
\end{corollary}

\begin{algorithm}
    \caption{Bayesian Privacy Filter}
    \label{algo:bayesian_privacy_filter}
  \begin{algorithmic}[1]
    \State  \textbf{GLOBAL STATE}:  privacy budget $\epsilon_g$, Data Universe $\mathcal{X}$
    \State \textbf{GLOBAL STATE}: map $P: \mathcal{X} \rightarrow \mathbb{R}$, where  $\forall x \in \mathcal{X}$,  $P(x) = 1$ when initialized.

    \INPUT query $M$
    \For{$y \in \text{Range}(M)$}
        \State Set $P' = P$.
        \State \label{step:times_Pr(y|x)} $ \forall x \in \mathcal{X}, $ Set $P'(x) = P'(x) \times Pr(M(x) = y) $.
        \State $L = \max P'(x) / \min P'(x)$ // realized privacy loss
        \If{$ L  > e^{\epsilon_g}$}
            \State \textbf{return} REJECT
        \EndIf
    \EndFor

    \State Execute $M$, and receive the result $y$.
    \State $ \forall x \in \mathcal{X}$, Set $P(x) = P(x) \times Pr(M(x) = y) $.
    \State \textbf{return} ACCEPT
        
  \end{algorithmic}
\end{algorithm}

Therefore, we can design a privacy filter that continuously accepts queries as long as Corollary \ref{cor:privacy_filter_accept} holds true for every query it receives. Such privacy filter is called Bayesian Privacy Filter:

 \begin{definition}
     A Bayesian Privacy Filter is a privacy filter using Eq. \eqref{eq:privacy_filter_accept} to accept or reject a query (Algorithm \ref{algo:bayesian_privacy_filter}).
 \end{definition}

\noindent The LDP guarantee has been proved in Corollary \ref{cor:privacy_filter_accept}. This algorithm maintains a map $P$, which records $\prod\limits_{1 \le i \le n} Pr(M_i(x) = y_i)$ for every $x$, accepted queries $M_i$, and their results $y_i$. For convenience, $M_i$ and $y_i$ are ignored, and $\prod\limits_{1 \le i \le n} Pr(M_i(x) = y_i)$ is denoted as $P(x)$, which will also be referred to as a likelihood function. The purpose of the algorithm is to ensure $\frac{\max_x P(x)}{\min_x P(x)} \le e^{\epsilon_g}$. 

Below is an example utilizing a Bayesian Privacy Filter:
\begin{example}
\label{ex_3}
    (Running Example). An object $X$ has 11 possible values as 0, 1, 2, ..., 10, and is protected by a Bayesian Privacy Filter with $\epsilon_g = 2\ln(\frac{3}{2})$ ($e^{\epsilon_g} = 2.25$). An analyst will query the values of $X$, $X^2$, ..., $X^i$, $X^{i+1}$, and so on, until the privacy filter rejects further queries. For any $i$, the LDP query against $X^i$ satisfies Eq. \eqref{eq:runing_example_prob}. Below is one possible outcome, where the table records all accepted queries:
    
    \begin{center}
    \begin{tabular}{ | c | c | c | }
    \hline
     $i$ & $y_i$ & $L(y_{\le i})$\\ 
     \hline
     1 & 1 & 1.50   \\  
     2 & 0 & 1.15  \\  
     3 & 1 & 1.50  \\  
     4 & 1 & 2.25  \\  
     \hline
    \end{tabular}
    \end{center}

    Since the realized privacy loss has reached $\epsilon_g$, all subsequent queries are rejected. 
\end{example}

The Bayesian Privacy Filter requires iteration over every potential output. However, if the iteration becomes too complex, there exists a simplified version:

\begin{definition}
    A Simplified Bayesian Privacy Filter is a privacy filter that accepts a query if and only if 
    \begin{equation}
        \log L(y_{\le n}) + \epsilon \le \epsilon_g , \notag
    \end{equation}
     where $y_{\le n}$ represents the accepted queries, and $\epsilon$ is the DP guarantee of a pending query.
\end{definition}

\noindent Further details are presented in Algorithm \ref{algo:simplified_bayesian_filter}.

\begin{algorithm}
    \caption{Simplified Bayesian Privacy Filter}
    \label{algo:simplified_bayesian_filter}
  \begin{algorithmic}[1]
    \State  \textbf{GLOBAL STATE}:  privacy budget $\epsilon_g$, Data Universe $\mathcal{X}$
    \State \textbf{GLOBAL STATE}: map $P: \mathcal{X} \rightarrow \mathbb{R}$, where  $\forall x \in \mathcal{X}$,  $P(x) = 1$ when initialized.

    \INPUT query $M$
    \State Read the DP parameter $\epsilon$ of $M$.
    \State $L = \max P(x) / \min P(x)$ // realized privacy loss
    \If{$L \: e^{\epsilon} > e^{\epsilon_g}$} \label{step:simplified_filter_check}
        \State  \textbf{return} REJECT
    \EndIf
    
    \State Execute $M$, and receive the result $y$.
    \State $ \forall x \in \mathcal{X}$, Set $P(x) = P(x) \times Pr(M(x) = y | x) $.
    \State \textbf{return} ACCEPT
        
  \end{algorithmic}
\end{algorithm}

\begin{theorem}
    A Simplified Bayesian Privacy Filter is $\epsilon_g$-LDP.
\end{theorem}

\noindent Proof: since the examined  query $M$ is $\epsilon$-LDP, the inequality

\begin{multline}
        \frac{\max\limits_{x \in \mathcal{X}} \prod\limits_{1 \le i \le n} Pr(M_i(x) = y_i) \times Pr(M(x) = y)}{\min\limits_{x' \in \mathcal{X}} \prod\limits_{1 \le i \le n} Pr(M_i(x') = y_i) \times Pr(M(x') = y)} \\ \le \frac{\max\limits_{x \in \mathcal{X}} \prod\limits_{1 \le i \le n} Pr(M_i(x) = y_i) }{\min\limits_{x' \in \mathcal{X}} \prod\limits_{1 \le i \le n} Pr(M_i(x') = y_i)} e^{\epsilon} 
    \label{eq:simplified_filter_proof_step_1}
\end{multline}

\noindent holds true. The right-hand side of Eq. \eqref{eq:simplified_filter_proof_step_1} corresponds to $P_{max} e^{\epsilon}$ in Step \ref{step:simplified_filter_check} of Algorithm \ref{algo:simplified_bayesian_filter}. Thus, the condition of Step \ref{step:simplified_filter_check} ensures
\begin{equation}
    \eqref{eq:simplified_filter_proof_step_1} \le e^{\epsilon_g}  , \notag
\end{equation}

\noindent which is equivalent to Corollary \ref{cor:privacy_filter_accept}. $\qed$

\subsection{Privacy Odometer}
\label{sec:odometer}
A privacy odometer dynamically tracks the privacy loss associated with accepted queries during a composition. An effective odometer should initialize at zero and must not exceed the privacy budget of the protected object. The remaining privacy budget (privacy budget - privacy loss) should be proportional to the capacity to accept new queries. The realized privacy loss has been shown to fulfill this role (see Section \ref{sec:privacy_filter}). Thus, our privacy odometer is defined as

\begin{definition}
    Bayesian Privacy Odometer is a function mapping from the accepted queries $(M_1, y_1, ..., M_j, y_j)$ after step \ref{step:interaction_accept_query} of the interaction between an adversary and an object (Algorithm \ref{algo:interaction}) to
    \begin{equation}
    \epsilon(y_{\le j}) = \log L(y_{\le j}) = \log \frac{\max\limits_{x \in \mathcal{X}} \prod\limits_{1 \le i \le j} Pr(M_i(x) = y_i)}{\min\limits_{x' \in \mathcal{X}} \prod\limits_{1 \le i \le j} Pr(M_i(x')=y_i)} .
    \notag
\end{equation} 
\end{definition}

Based on Eq. \eqref{eq:privacy_filter_dp}, $\epsilon(y_{\le j})$ will never exceed $\epsilon_g$ if the interaction is managed by a Bayesian Privacy Filter. When $\epsilon(y_{\le j})$ equals $\epsilon_g$, no new queries with a non-zero DP guarantee can be accepted, as specified by Eq. \eqref{eq:privacy_filter_accept}. Furthermore, if a Simplified Bayesian Privacy Filter is used, only queries with a DP guarantee $\epsilon$ that does not exceed $\epsilon_g - \epsilon(y_{\le j})$ will be accepted. 

Another observation is that $\epsilon(y_{\le j})$ could decrease. Define $x_1$ and $x_2$ to be $\argmax \prod\limits_{1 \le i \le j} Pr(M_i(x) = y_i) $ and $\argmin \prod\limits_{1 \le i \le j} Pr(M_i(x) = y_i) $, respectively. If the result $y$ of the next query satisfies $Pr(y | x_1) < Pr(y | x_2)$, $\epsilon(y_{\le j})$ will decrease.

\subsection{Extend to Approximate Differential Privacy}
\label{sec:aldp}
Local Approximate Differential Privacy, denoted as $(\epsilon, \delta)$-LDP, is defined as follows: 

\begin{definition}
   A query $M$ is $(\epsilon, \delta)$-LDP, if and only if, for any $x, x' \in \mathcal{X}$ and any subset $S$ of the outputs generated by $M$,
\begin{equation}
    Pr(M(x) \in S) \le e^{\epsilon} Pr(M(x') \in S) + \delta . \notag
\end{equation} 
\end{definition}

\noindent This definition is equivalent to:

\begin{definition}
\label{def:eqv_ed_ldp}
   For any $x \in \mathcal{X}$ and with the probability of $1 - \delta$, an output $y$ generated by $M(x)$ satisfies the realized privacy loss $L(y) < e^{\epsilon}$ \cite{kasiviswanathan2014semantics}.
\end{definition}

Previous research has introduced a category of privacy filters designed to enforce $(\epsilon, \delta)$-LDP for fully adaptive composition \cite{rogers2016privacy, whitehouse2022fully}. These privacy filters either accept a query or halt the composition. When the composition is halted, these filters guarantee, with the probability of $1 - \delta$, that the composition will have $L(\mathbf{y}) \le e^\epsilon$, irrespective of the outputs $\mathbf{y}$ of the accepted queries. Distinguished from Bayesian Privacy Filters, these privacy filters are termed output-independent filters. Although derived for CDP, they are applicable to LDP due to the equivalence of fully adaptive composition in both LDP and CDP. An $(\epsilon_g, \delta_g)$-LDP output-independent filter can be enhanced to an $(\epsilon_g, \delta_g)$-Bayesian Privacy Filter in the following manner: consider the situation where an output-independent filter decides to halt the composition. With the probability of $\delta_g$, $L(\mathbf{y})$ will exceed $e^{\epsilon_g}$. In this case, as the $\epsilon_g$-LDP guarantee has already been violated, and there is no further constraint for $L(\mathbf{y})$, $L(\mathbf{y})$ can be infinitely large. Consequently, the Bayesian Privacy Filter is permitted to continue accepting an infinite number of arbitrary queries, even directly outputting the true value $x$, as allowed under Definition $\ref{def:eqv_ed_ldp}$. However, in good faith, the Bayesian Privacy Filter chooses to terminate the composition to prevent further privacy leaks. Conversely, with the probability of $1 - \delta_g$, $L(\mathbf{y})$ remains below $e^{\epsilon_g}$, and the Bayesian Privacy Filter is obligated to sustain this. Thus, the filter can accept queries as long as the $L(\mathbf{y})$ after acceptance does not surpass $\epsilon_g$. Detailed implementation is provided in Algorithm \ref{algo:approximate_filter}.

\begin{algorithm}
    \caption{Bayesian Privacy Filter for $(\epsilon_g, \delta_g)$-LDP}
    \label{algo:approximate_filter}
  \begin{algorithmic}[1]
    \State  \textbf{GLOBAL STATE}:  privacy budget $(\epsilon_g, \delta_g)$
    \State  \textbf{GLOBAL STATE}:  output-independent $(\epsilon_g, \delta_g)$-LDP Privacy Filter $F_0$
    \State  \textbf{GLOBAL STATE}:  $\epsilon_g$-LDP Bayesian Privacy Filter $F_B$
    \State  \textbf{GLOBAL STATE}: the outputs of the currently accepted queries $\mathbf{y}$.

    \INPUT query $M$
    \If{$F_0$ is not halted}
        \State $F_0$ receives $M$ and outputs ACCEPT or HALT
        \If{$F_0$ outputs ACCEPT}
            \State Execute $M$ and add the output into $\mathbf{y}$
            \State\Return ACCEPT
        \Else
            \State $F_0$ becomes halted.
        \EndIf
    \EndIf
    \If{$L(\mathbf{y}) > \epsilon_g$}
        \State\Return REJECT
    \EndIf
    \State // $F_B$ can accept $M$ only if $L(\mathbf{y}) \le \epsilon_g$ given any output from $M$.
    \If{$F_B$ can accept $M$} 
        \State $F_B$ accepts, executes $M$, and add the output into $\mathbf{y}$
        \State\Return ACCEPT
    \EndIf
    \State\Return REJECT
  \end{algorithmic}
\end{algorithm}

The LDP guarantee of an $(\epsilon_g, \delta_g)$-Bayesian Privacy Filter is explained above; therefore, no formal proof is provided here. Despite the existence of the $(\epsilon_g, \delta_g)$-Bayesian Privacy Filter, the subsequent sections of the article will concentrate exclusively on the $\epsilon_g$-Bayesian Privacy Filter.

\subsection{Role of Bayesian Belief}
\label{sec:role_of_bayesian_belief}
This paper interprets the rationale behind a Bayesian Privacy Filter as ensuring the realized privacy loss of an adaptive composition to not exceed the global LDP guarantee. The semantics of the realized privacy loss stem from the adoption of Bayesian belief. The existence of the Bayesian belief is supported by either (1) the presence of a prior distribution of the studied object or (2) an adversary adopting a subjective belief. Notably, the existence of the Bayesian belief is not mathematically required by DP\cite{tschantz2017differential}. However, Theorem \ref{thm:privacy_filter_ldp} is proven without reliance on Bayesian belief, and its correctness does not depend on adopting such a framework. For individuals who do not subscribe to Bayesian belief, the LDP guarantee of a Bayesian Privacy Filter remains valid, though they may not accept the semantic interpretation of realized privacy loss.

\subsection{Application to QIF}
\label{sec:application_to_QIF}
As stated in Theorem \ref{thm:realized_g_leakage_ub}, the realized privacy loss serves as an upper bound on any realized $g$-leakage. Since a Bayesian Privacy Filter ensures that the realized privacy loss never exceeds $e^{\epsilon}$, it also guarantees that, for any $g$ and prior belief $Q$, the realized $g$-leakage of the composed queries will not exceed $e^{\epsilon}$. Similarly, the output of a Bayesian Privacy Odometer also provides an upper bound on any realized $g$-leakage of the composed queries.

\section{Algorithms to Compute Realized Privacy Loss}
\label{sec:algo_to_compose}

The Bayesian Privacy Filter and Odometer calculate a realized privacy loss (left hand side of Eq. \eqref{eq:privacy_filter_dp}) after receiving a query. The initial implementation (see Algorithm \ref{algo:bayesian_privacy_filter} and \ref{algo:simplified_bayesian_filter}) relies on maintaining the likelihood function $P(x)$ for every $x \in \mathcal{X}$ to compute the realized privacy loss. However, updating $P(x)$ involves iterating over every $x$. In cases where $|\mathcal{X}|$ is excessively large, such as when $\mathcal{X}$ is a continuous value, computing $P(x)$ becomes impractical. In this section, we will introduce a method to numerically compute the realized privacy loss for an object with multi-dimensional continuous values without the need to maintain $P(x)$ for every $x$.

\subsection{Upper Bound of Realized Privacy Loss}

Computing a realized privacy loss requires both maximum and minimum of the likelihood function $P(x)$. While determining the exact extrema can be computationally expensive, it's often more feasible to calculate the lower bound $\min_x P(x)$ and upper bound $\max_x P(x)$. Let $LB \le \min_x P(x)$ and $\max_x P(x) \le UB$. The realized privacy loss can then be expressed as:

\begin{equation}
    L(y) = \frac{\max\limits_x P(x)}{\min\limits_{x} P(x)} \le \frac{UB}{LB} .
    \label{eq:L_bound}
\end{equation}

Bayesian Privacy Filter and Odometer may instead utilize the upper bound of $L(y)$ for privacy accounting, and their privacy guarantee remains valid. The subsequent section will introduce a method to determine $LB$ and $UB$ for $P(x)$ using the technique of Difference of Convex Programming.

\subsection{Difference of Convex Programming}
\label{sec:DC}

Difference of Convex (DC) Programming \cite{tao1997convex} with box constraints addresses the following problem:

\begin{equation}
    \begin{gathered}
            \min f(x) - g(x) \\
    \text{subject to  } x_{i, lb} \le x_i \le x_{i, ub}, \;\; 1 \le i \le m
    \end{gathered}
    \label{eq:DC}
\end{equation}

\noindent where $x \in \mathbb{R}^m$; $f(x)$ and $g(x)$ are convex functions. Despite the objective being the minimization of $f(x) - g(x)$, the corresponding maximization can be obtained by interchanging $f(x)$ and $g(x)$. The lower bound of the global solution can be determined by relaxing $g(x)$ into an affine (linear) function $h(x)$ under the conditions that (1) $g(x) \le h(x)$ within a compact (closed and bounded) set of $x$, and (2)  $h(x) \rightarrow g(x)$ as the compact set of $x$ approaches infinitesimally small. This results in a convex optimization problem, with $\min f(x) - h(x)$ serving as the lower bound for $\min f(x) - g(x)$. The function $h(x)$ is referred to as the concave envelope of $g(x)$. Utilizing the branch-and-bound method (Section \ref{sec:bnb}), the lower bound of $\min f(x) - g(x)$, i,e., $LB$ in Eq. \eqref{eq:L_bound}, will converge gradually to $\min_{x} P(x)$. $UB$ can be acquired similarly with $\min g(x) - f(x)$.

Notably, $\log P(x)$ can be expressed as

\begin{equation}
    \log P(x) = \sum\limits_{1 \le i \le n} \log P_i(x) ,
    \label{eq:sum_log_P}
\end{equation}

\noindent where $P_i(x) = Pr(M_i(x) = y_i | x)$. If $\log P_i(x)$ can be expressed as the difference of two convex functions, given that the sum of convex functions is still convex, minimizing/maximizing $\log P(x)$ transforms into a DC programming.

\subsection{Translate Queries into Difference of Convex Functions}
\label{sec:query_to_dc}

This section introduces three types of LDP queries with likelihood functions $P(x)$ that can be represented in the form of DC programming. This assumes that their LDP is enforced using the perturbation algorithm outlined in Section \ref{sec:LDP_of_continuous_value}.

\subsubsection{Linear Regression}

Consider a vector $x$, and let $y$ be the result of a linear regression against it, with the form $y = \theta x + c$, and $y \in [a, b]$ under the constraints of $x$. By replacing $y$ in Eq. \eqref{eq:linear_pr} with $\theta x + c$, the conditional probability of an output given $x$ is:

\begin{equation}
    Pr(b | x) =  \frac{1}{b - a} \frac{ e^{\epsilon} - 1}{ e^{\epsilon} + 1} (\theta x + c - a) +  \frac{1}{ e^{\epsilon} + 1} , \notag
\end{equation}

\noindent and $Pr(a | x) = 1 - Pr(b | x)$. It is evident that both $Pr(b|x)$ and $Pr(a|x)$ are linear functions of $x$. For simplicity, we express the conditional probability as:

\begin{equation}
    Pr(o|x) = \alpha x + \beta \notag
\end{equation}

\noindent where $o$ is either $a$ or $b$, and $\alpha$ and $\beta$ represent the slope and intercept of the linear relationship. The logarithmic form $\log Pr(o|x)$ is concave, allowing it to be incorporated into DC programming, where $\log Pr(o|x)$ serves as a component of $-g(x)$.

\subsubsection{Truncated Linear Regression} 

Similar to the linear regression, the value $y$ in truncated linear regression is represented as:

\begin{equation}
    y = \min\{b, \max\{a, \theta x + c\}\} . \notag
\end{equation}

\noindent After enforcing LDP, the conditional probability of an output is given by
\begin{multline}
    Pr(b | x) = \min\{ \frac{e^{\epsilon}}{ e^{\epsilon} + 1}, \max\{\frac{1}{ e^{\epsilon} + 1}, \\ \frac{1}{b - a} \frac{ e^{\epsilon} - 1}{ e^{\epsilon} + 1} (\theta x + c - a) +  \frac{1}{ e^{\epsilon} + 1} \}\} , \notag
\end{multline}

\noindent and $Pr(a | x) = 1 - Pr(b | x)$. For simplicity, the conditional probability is rewritten as: 
\begin{equation}
    Pr(o|x) = \min\{\frac{e^{\epsilon}}{ e^{\epsilon} + 1}, \max\{\frac{1}{ e^{\epsilon} + 1},  \alpha x + \beta \}\} . \notag
\end{equation} 

\noindent $\log Pr(o|x)$ can be expressed as $f(x) - g(x)$, where:

\begin{equation}
    f(x) = \log(\frac{1}{ e^{\epsilon} + 1}) - \log(\min\{\frac{1}{ e^{\epsilon} + 1}, \: \alpha x + \beta\}) , \notag
\end{equation} 

\noindent and 
\begin{equation}
    g(x) = - \log(\min\{\frac{e^{\epsilon}}{ e^{\epsilon} + 1}, \: \alpha x + \beta\})  . \notag
\end{equation} 

\noindent Both $f(x)$ and $g(x)$ are convex (proof is omitted), demonstrating that $\log Pr(o|x)$ can be accommodated within DC programming.

\subsubsection{Logistic Regression}
\label{sec:logistic_regression}

Let $x$ be a vector, and denote $y$ as the result of logistic regression:

\begin{equation}
    y = \frac{e^{\theta x + c}}{e^{\theta x + c} + 1} . \notag
\end{equation}

\noindent If LDP is enforced for $y$ using the perturbation algorithm in Section \ref{sec:LDP_of_continuous_value}, then the algorithm will output either 0 or 1. Substituting the $y$ into Eq. \eqref{eq:linear_pr}, the conditional probability of an output given $x$ becomes:

\begin{equation}
    Pr(1 | x) = \frac{ e^{\epsilon} - 1}{ e^{\epsilon} + 1} \frac{e^{\theta x + c}}{e^{\theta x + c} + 1} +  \frac{1}{ e^{\epsilon} + 1} \notag
\end{equation}

\noindent and $Pr(0 | x) = 1 - Pr(1 | x)$. For simplicity, the formula can be rewritten as:

\begin{equation}
    Pr(o|x) = \frac{\alpha e^{\theta x + c} + \beta}{e^{\theta x + c} + 1} , \notag
\end{equation}

\noindent where $o$ is either 0 or 1, and $\alpha$ and $\beta$ are the parameters after the refactor. Defining

\begin{equation}
    f(x) = \log(\alpha e^{\theta x + c} + \beta)  \notag
\end{equation}
\noindent and 
\begin{equation}
    g(x) = \log(e^{\theta x + c} + 1)  , \notag
\end{equation}

\noindent we have $\log Pr(o|x) = f(x) - g(x)$. Given $f(x)$ and $g(x)$ are both convex (because $\log(e^x + 1)$ is convex and $\theta x + c$ is a linear transformation), $\log Pr(o|x)$ can be incorporated into DC programming accordingly.

\section{Evaluation of Bayesian Privacy Filter}
\label{sec:evaluation_of_bayesian_composition}

In this section, we will use the term \textit{Bayesian} composition to represent the interaction between an adversary and an object managed by a Bayesian Privacy Filter, as compared to the basic composition. It is important to note that the adversary is assumed to be rational and greedy. This implies that the adversary in the Bayesian composition will not halt the composition if the Bayesian Privacy Filter can still accept any queries the adversary likes to send. We will delve into efficiency experiments for Bayesian composition, wherein a Bayesian Privacy Filter receives a stream of LDP queries until the filter begins rejecting queries. The pseudocode is presented in Algorithm \ref{algo:bayesian_experiment}. Eventually, we discuss the use case of Bayesian composition.

\begin{algorithm}
    \caption{Bayesian Composition Experiment}
    \label{algo:bayesian_experiment}
  \begin{algorithmic}[1]
    \INPUT a stream of queries $\mathbf{q}$
    \State an object $x$ protected by Bayesian Privacy Filter $F$
    \State $O = []$ // outputs
    \Loop
        \State next query $q$ from $\mathbf{q}$
        \State $F$ receives $q$ and returns RESULT
        \If{RESULT is REJECT} 
            \State \Return $O$
        \EndIf
        \State $F$ executes $q$ with output $o$.
        \State Append $(q, o)$ to $O$
    \EndLoop
  \end{algorithmic}
\end{algorithm}

\subsection{Composing Linear Regressions}
\label{sec:compsing_linear_regression}

\begin{figure*}[ht!]
\centering
\subfloat[]{\label{fig:regression-a} \includegraphics[width=0.24\textwidth]{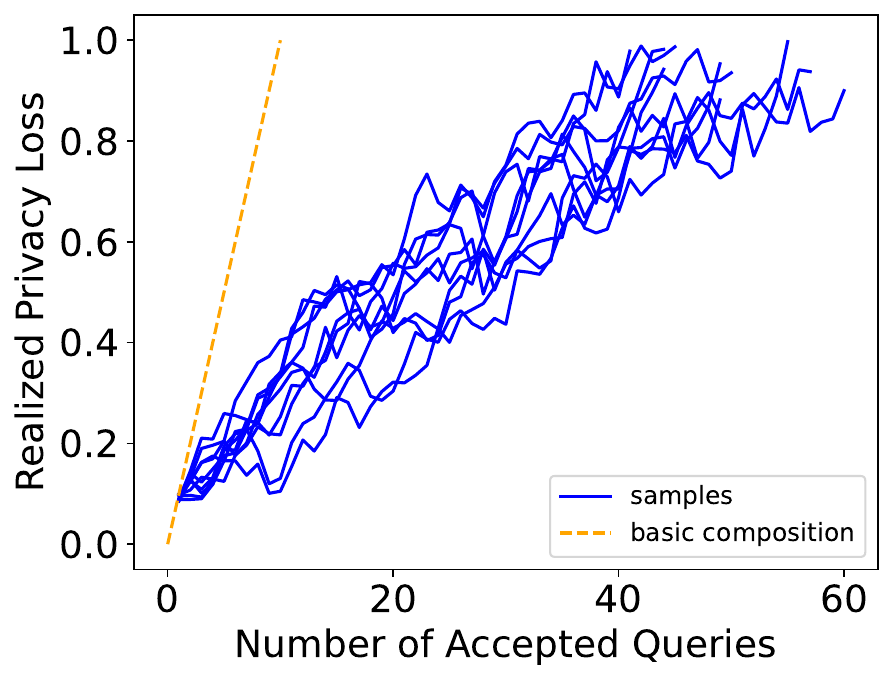}}
\hfill
\subfloat[]{\label{fig:regression-b} \includegraphics[width=0.24\textwidth]{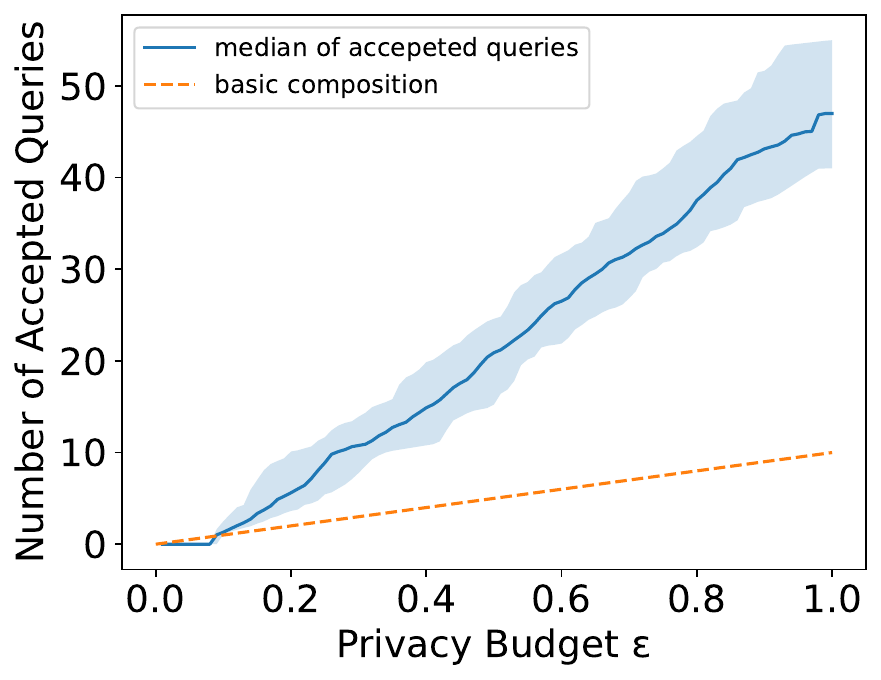}}%
\hfill
\subfloat[]{\label{fig:regression-c} \includegraphics[width=0.24\textwidth]{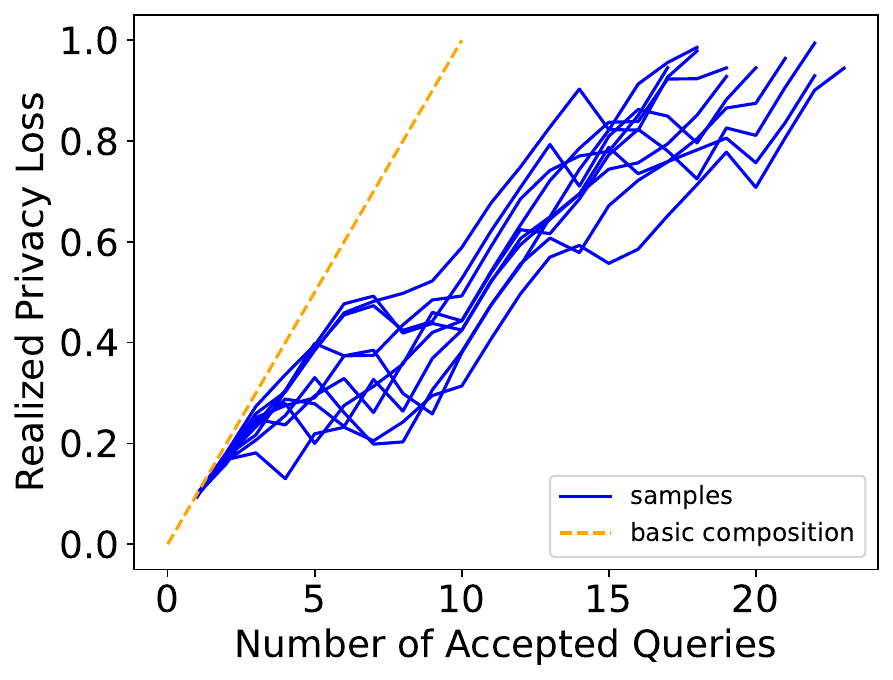}}%
\hfill
\subfloat[]{\label{fig:regression-d} \includegraphics[width=0.24\textwidth]{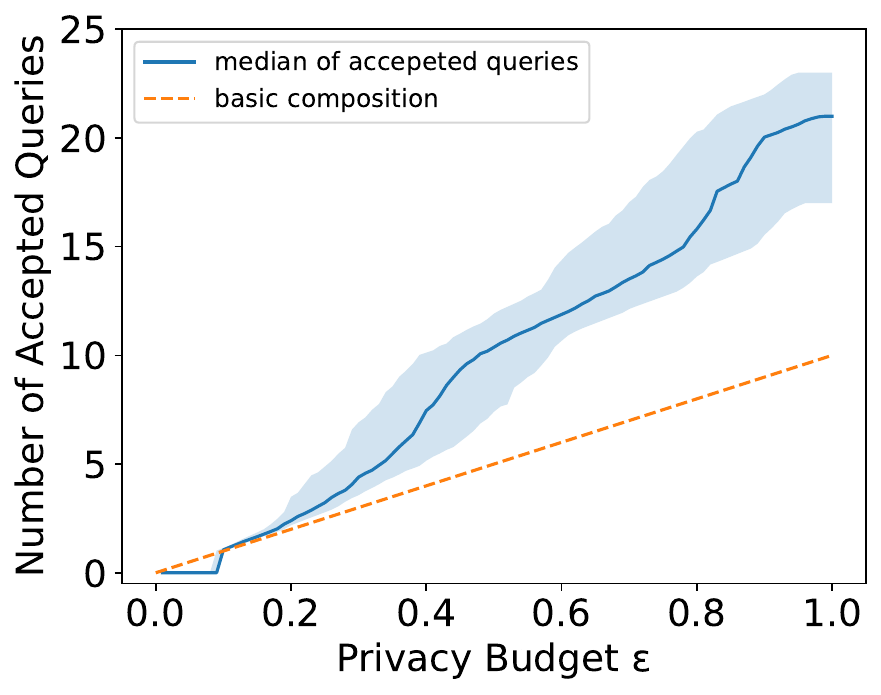}}%
\caption{Results for Composition Experiments. (a) realized privacy loss vs. number of accepted queries for 10 sampled linear regressions; (b) confidential interval (10th to 90th percentile, same for (d)) of number of accepted queries vs. privacy budget for linear regressions; (c) realized privacy loss vs. number of accepted queries for 10 sampled logistic regressions; and (d) confidential interval of number of accepted queries vs. privacy budget for logistic regressions.}
\label{fig:regression}
\end{figure*}

An experiment is designed to continually send linear regressions to query an object protected by a Bayesian Privacy Filter until the filter starts rejecting. This section will compare the number of the queries allowed by the privacy filter with that by the basic composition. The object has a 9-dimensional domain and each dimension ranges from $[-1, 1]$, denoted as $x \in [-1, 1]^9$. For convenience, the true value $x$ is set to be $\mathbf{0}$ for all dimensions, unknown to the analyst querying the object. The object is assigned a privacy budget $\epsilon = 1.0$. Each linear regression, with 10 parameters denoted as $\mathbf{\theta} = \{\theta_0, \theta_1, ..., \theta_9\}$, has $\theta_i$ uniformly sampled between [-1, 1]. The $\mathbf{\theta}$ of each query is normalized to be  $|| \mathbf{\theta} ||_1 = 1$ ($L_1$ Norm). Before enforcing LDP, the output of the linear regression against $x$ is $y = \theta_0 + \sum_{1 \le i \le 9} \theta_i x_i$, and $y \in [-1, 1]$. LDP is enforced against $y$, following Section \ref{sec:LDP_of_continuous_value}, with a guarantee of $\epsilon = 0.1$ per query. The privacy filter groups every 10 regressions to reduce computation (see Section \ref{sec:group_queries}), and the realized privacy loss is computed using the branch-and-bound method (Section \ref{sec:bnb}).

The experiment follows Algorithm \ref{algo:bayesian_experiment} and has been executed 50 times. Fig. \ref{fig:regression-a} randomly samples 10 runs, illustrating that the realized privacy loss increases with the number of queries accepted by the privacy filter. In all sampled experiments, the realized privacy loss grows slower than that of basic composition.

Additionally, the confidence interval of the number of accepted queries given a privacy budget $\epsilon$ is of interest. This refers to the number of queries accepted right before the privacy budget is exhausted. Their 10th percentile (lower bound), median, and 90th percentile (upper bound) are computed as Fig \ref{fig:regression-b}. The results indicate that the number of accepted queries increases approximately linearly with $\epsilon$. The median of accepted queries at $\epsilon = 1.0$ is 47, while basic composition only accepts 10 queries.

\subsection{Composing Logistic Regressions}

An experiment akin to Section \ref{sec:compsing_linear_regression} is conducted for logistic regressions. All the setups remain identical except for the parameter $\mathbf{\theta}$ of each logistic regression. Like linear regression, $\mathbf{\theta}$ has 10 dimensions, but each dimension is uniformly sampled between $[-10, 10]$ without normalization. Before enforcing LDP, the output $y$ is $\frac{1}{1 + e^{-(\theta_0 + \sum \theta_i x_i)}}$, ranging from 0 to 1. LDP is enforced against $y$ following Section \ref{sec:LDP_of_continuous_value}. 

The experiment is run for 50 times.  Fig. \ref{fig:regression-c} randomly samples 10 runs, illustrating that the realized privacy loss increases with the number of queries accepted by the Bayesian Privacy Filter. Similar to the linear regression experiment, all sampled experiments exhibit a realized privacy loss growing slower than that of basic composition.

Also, akin to Section \ref{sec:LDP_of_continuous_value}, the confidence interval of the number of accepted logistic regressions given a privacy budget $\epsilon$ is plotted in Fig. \ref{fig:regression-d}, with upper and lower bounds as the 90th and 10th percentiles, respectively. Similar to linear regression, the number of accepted logistic regressions increases approximately linearly with $\epsilon$. The median of accepted queries at $\epsilon = 1.0$ is 21, while for linear regression, it is 47. Nevertheless, it still outperforms basic composition.

\subsection{Read-world Example: Compose Regressions of Healthcare data}
This section describes how a Bayesian Privacy Filter preserves the privacy budget when processing real-world queries. An analyst is interested in a group's health checkup data, represented as follows with mock data:

\begin{center}
    \begin{tabular}{ | c | c | c | c | c | c  }
    \hline
     ID & age & sex & high blood pressure & BMI & ... \\ 
     \hline
     0 & 34 & F & 110 & 18 &   \\  
     1 & 52 & M & 140 & 23 &   \\  
     ... \\ 
    \end{tabular}
    \end{center}

\noindent The analyst trains three logistic regressions from public datasets \cite{heart, stroke, diabetes} to predict the likelihood of individuals having heart disease, stroke, and diabetes. The mean of these probabilities represents the disease incidence rates for the group. They also train a linear regression from a public dataset \cite{sleep} to predict individuals' sleep duration, and the mean of the predicted values yields the group's average sleep duration. The regression parameters are provided in Appendix \ref{sec:privacy_loss_of_health_data}. The table below displays the query results for the mock data before applying LDP:
\begin{center}
    \begin{tabular}{ | c | c | c | c | c |   }
    \hline
     ID & $Pr_{\text{heart disease}}$ & $Pr_{\text{stroke}}$ & $Pr_{\text{diabetes}}$ & sleep duration \\ 
     \hline
     0 & 0.94 & 0.06 & 0.094 & 7.50   \\  
     1 & 0.45 & 0.027 & 0.24 & 7.32   \\  
     ... \\ 
    \end{tabular}
    \end{center}

\noindent To enforce LDP, these regressions are discretized and perturbed following Section \ref{sec:LDP_of_continuous_value}. Each individual in the group is assigned a privacy budget of $\epsilon = 4$, with each regression having an DP guarantee of $\epsilon=1$. Their realized privacy loss is computed using the branch-and-bound method (Section \ref{sec:bnb}). As each regression randomly returns two discrete values, there are 16 possible outcomes. The realized privacy losses are detailed in Appendix \ref{sec:privacy_loss_of_health_data}. In the worst case, 3.7 out of 4 of the privacy budget $\epsilon$ is consumed, while the median privacy loss is only 2.46 out of 4, indicating that at least half of the group can receive additional queries with an DP guarantee $\epsilon$ of at least 1.54.

\section{Use Case of Bayesian Composition}

Bayesian composition is at least as good as basic composition, because basic composition is equivalent to considering the likelihood function of each composed queries independently in Bayesian composition. Thus, the Bayesian composition can seamlessly replace the basic composition wherever it is employed, especially in privacy accounting. 

However, the necessity of the replacement is worth considering, because Bayesian composition requires computing realized privacy loss, which might be expensive in some cases. For instance, when querying an object with continuous values, the branch-and-bound method (Section \ref{sec:bnb}) is required, which is generally computation-intensive. The evaluation in Section \ref{sec:evaluation_of_bayesian_composition} highlights that Bayesian composition of linear and logistic regressions against objects with continuous values can be 2-5 times as efficient as basic composition. In scenarios where privacy budgets are scarce, deploying Bayesian composition becomes a viable and worthwhile option.




\section{Conclusion}

\begin{enumerate}
\item DP guarantees could overestimate the actual knowledge gained by adversary after observing a query.

\item The maximum change in belief regarding any statement specific to a queried object has been derived and is referred to as the realized privacy loss.

\item The Bayesian Privacy Filter fully utilizes the privacy budget of an adaptive composition by accepting queries if the realized privacy loss does not exceed the DP guarantee. Using the realized privacy loss as a privacy odometer accurately represents the capacity to accept new queries.

\item A branch-and-bound method is devised to compute the realized privacy loss of linear and logistic regressions against objects with continuous values.

\item The Bayesian composition of linear and logistic regressions is observed to accept 2-5 times as many queries as basic composition.
\end{enumerate}

\bibliographystyle{IEEEtran}
\bibliography{references}  
%

\appendices

\section{Proof of the Equivalence between LDP and a Global Learning Limit}
\label{sec:proof_LDP=learning_limit}
\subsection{Proof of Lemma \ref{lem:bound_of_Q_f}}

We will begin with the first inequality. Here, $Pr(M(x)= y)$ is abbreviated as $Pr(y | x)$. For any statement $f$ specific to $X$, after a query $M(X)$ yielding a result $y$, we have
\begin{equation}
    \frac{Q(f | y)}{Q(f)} = \frac{\sum\limits_{x \in \mathcal{X}} Pr(y | x) Pr(f(x)) Q(x)} {\sum\limits_{x' \in \mathcal{X}} Pr(y | x') Q(x') \sum\limits_{x \in \mathcal{X}} Pr(f(x)) Q(x)} .
    \label{eq:knowledge_bound_step_1}
\end{equation}

\noindent Given $\sum\limits_{x' \in \mathcal{X}}Q(x') = 1$, we have $\sum\limits_{x' \in \mathcal{X}} Pr(y | x') Q(x') \le \max\limits_{x' \in \mathcal{X}}Pr(y | x')$. Additionally, we have $\sum\limits_{x \in \mathcal{X}} Pr(y | x) Pr( \allowbreak f(x))   Q(x) \ge \min\limits_{x \in \mathcal{X}}Pr(y | x) \sum\limits_{x \in \mathcal{X}} Pr(f(x)) Q(x)$. Thus, 

\begin{equation}
     \eqref{eq:knowledge_bound_step_1} \ge \frac{\min\limits_{x \in \mathcal{X}}Pr(y | x) \sum\limits_{x \in \mathcal{X}} Pr(f(x)) Q(x)}{\max\limits_{x' \in \mathcal{X}}Pr(y | x') \sum\limits_{x \in \mathcal{X}} Pr(f(x)) Q(x)} \ge \frac{\min\limits_{x \in \mathcal{X}}Pr(y | x)}{\max\limits_{x' \in \mathcal{X}}Pr(y | x')} .
    \label{eq:knowledge_bound_step_2}
\end{equation}

\noindent The proof of the first inequality is complete. The second inequality can be proven in a similar manner by scaling in the opposite direction, and the details are omitted here. $\qed$.

\subsection{Proof of Theorem \ref{thm:ldp_to_learnig_limit}}

Given Lemma \ref{lem:bound_of_Q_f}, if $M$ is $\epsilon$-LDP, from Eq. \ref{eq:knowledge_bound_step_2}, we obtain $\forall f, y$:
\begin{equation}
    \eqref{eq:knowledge_bound_step_1} \ge \frac{\min\limits_{x \in \mathcal{X}}Pr(y | x)}{\max\limits_{x' \in \mathcal{X}}Pr(y | x')} \ge e^{-\epsilon} . \notag
\end{equation}

\noindent The above inequality is equivalent to the first inequality in Eq. \eqref{eq:learning_limit}. Similarly, we have Eq. $\eqref{eq:knowledge_bound_step_1} \le e^{\epsilon}$, which is equivalent to the second inequality in Eq. \eqref{eq:learning_limit}. Thus, $M$ provides an $\epsilon$-learning limit for all statements specific to $X$. $\qed$.

\subsection{Proof of Lemma \ref{lem:Q_f_bound_eq}}

Lemma \ref{lem:Q_f_bound_eq} is equivalently stated as follows: for any $\delta > 0$ and any $y$, there exists a statement $f$ and a prior belief $Q$, the inequality
    \begin{equation}
        \frac{Q(f | y)}{Q(f)} \ge \frac{\max\limits_{x \in \mathcal{X}}Pr(M(x) = y)}{\min\limits_{x' \in \mathcal{X}}Pr(M(x') = y)} -\delta
        \label{eq:proof_Q_f_bound_eq}
    \end{equation}
    holds.

We only focus on  $\delta < \frac{\max\limits_{x \in \mathcal{X}}Pr(M(x) = y)}{\min\limits_{x' \in \mathcal{X}}Pr(M(x') = y)}$. Otherwise, the proof is trivial. Given any $y$, define a statement $f$ satisfying $Pr(f(x)) = 1$ only if $x = \argmax_{x \in \mathcal{X}} Pr(y | x)$; otherwise $Pr(f(x)) = 0$. Also define a prior distribution $Q$ as follows:
\begin{equation}
        Q(x) = 
     \begin{cases}
        p, \text{if } x = \argmax_{x \in \mathcal{X}} Pr(y | x) \\
        1 - p, \text{if } x = \text{argmin}_{x \in \mathcal{X}} Pr(y | x) \\
        0, \text{ otherwise.} \\
    \end{cases}, \notag
\end{equation}

\noindent where $p \le \frac{\delta p_{min}^2}{(p_{max} - p_{min})(p_{max} - \delta p_{min})}$, $p_{max} = \max_{x \in \mathcal{X}} Pr(y | x)$, and  $p_{min} = \min_{x \in \mathcal{X}} Pr(y | x)$. Substituting $Q$ and $f$ into Eq. \eqref{eq:proof_Q_f_bound_eq}, we can see the inequality is satisfied. $\qed$

\subsection{Proof of Theorem \ref{thm:learning_limit_to_LDP}}

 We will instead prove the following contrapositive of the theorem:

\begin{proposition}
    If a query $M$ is not $\epsilon$-LDP, there exists a statement $f$ specific to $X$ for which $M(X)$ cannot provide an $\epsilon$-learning limit.
\end{proposition}

\noindent We also need the contrapositive of the definition of $\epsilon$-learning limit (Definition \ref{def:learning_limit}):

\begin{proposition}
    A query $M(X)$ cannot provide an $\epsilon$-learning bound for a statement $f$ specific to $X$, if and only if there exists an output $y$ from $M$ and a prior belief $Q$ such that
    \begin{equation}
         Q(f | y)  < e^{-\epsilon} Q(f) \notag
    \end{equation}
    or

    \begin{equation}
        Q(f | y) > e^{\epsilon} Q(f) . \label{eq:not_learning_limit_second}
    \end{equation}
\end{proposition}

\noindent Suppose $M$ is not $\epsilon$-LDP, and we have

\begin{equation}
        \max\limits_{x \in \mathcal{X}} Pr(y | x) > e^{\epsilon} \min\limits_{x \in \mathcal{X}}  Pr(y | x) . \notag
\end{equation}

\noindent Rewrite the above equation as $\max\limits_{x \in \mathcal{X}} Pr(y | x) = e^{\epsilon} \min\limits_{x \in \mathcal{X}}  Pr(y | x) + \theta $. Setting $\delta$ in Eq \eqref{eq:proof_Q_f_bound_eq} to be $\theta$, we have

\begin{equation}
    \frac{Q(f | y)}{Q(f)} \ge e^{\epsilon} + \frac{\theta}{\min\limits_{x \in \mathcal{X}}  Pr(y | x)} - \theta . \notag
\end{equation}

\noindent Notice that $\min\limits_{x \in \mathcal{X}}  Pr(y | x) < 1$. Otherwise, $\max\limits_{x \in \mathcal{X}}  Pr(y | x)$ will be larger than one. Therefore, $\frac{\theta}{\min\limits_{x \in \mathcal{X}}  Pr(y | x)} > \theta $, and we have

\begin{equation}
    \frac{Q(f | y)}{Q(f)} > e^{\epsilon} + \theta - \theta = e^{\epsilon}. \notag
\end{equation}

\noindent which is equivalent to Eq. \eqref{eq:not_learning_limit_second}. The satisfaction of Eq. \eqref{eq:not_learning_limit_second} concludes that the query $M$ cannot provide an $\epsilon$-learning limit for a statement $f$. Therefore, the contrapositive of Theorem \ref{thm:learning_limit_to_LDP} is proved, as is the theorem. $\qed$

\section{Proof of QIF-related Theorems}
\label{sec:proof_qif}

To prove Theorem \ref{thm:realized_g_leakage_ub}, we begin with
\begin{multline}
    LK_g(Q | y) = \frac{V_g(Q | y)}{V_g(Q)} \\ = \frac{ \max_w Q(f_{g, w} | y)}{ \max_w Q(f_{g, w})} \le \max_{ w}  \frac{ Q(f_{g, w} | y)}{ Q(f_{g, w})}.
    \notag
\end{multline}

\noindent Reformulate the second inequality of Lemma \ref{lem:bound_of_Q_f} and apply it to the max-case $g$-leakage: 
\begin{multline*}
    \forall y: \max_{g, Q} LK_g(Q | y) \le \max_{Q, g, w}  \frac{ Q(f_{g, w} | y)}{ Q(f_{g, w})} \le \\ \max_{Q, f}  \frac{ Q(f | y)}{ Q(f)} \stackrel{\text{Lem \ref{lem:bound_of_Q_f}}}{\le}  \frac{\max\limits_{x \in \mathcal{X}}Pr(M(x) = y | x)}{\min\limits_{x' \in \mathcal{X}}Pr(M(x') = y | x')} = L(y) . \qed
\end{multline*}

\noindent If a query is $\epsilon$-LDP, we have $\max_y L(y) \le e^{\epsilon}$, so
\begin{equation*}
    \max_{g, Q, y} LK_g(Q | y) \le \max_y L(y) \le e^{\epsilon} ,
\end{equation*}

\noindent which proves Corollary \ref{cor:max_case_g_leakage_ub}. $\qed$

\section{Grouping Queries}
\label{sec:group_queries}

This section introduces a method to break down the computation of the extrema of $P(x)$ into smaller problems, each of which computes the extrema of $P(x)$ for a subset of the accepted queries. It is particularly helpful when complexity of computing the extrema of $P(x)$ grows superlinearly (e.g. exponentially) with the number of composed queries. Given a sequence of queries, each denoted as $M_i$ with $y_i$ as the corresponding output, the upper bound of their likelihood function $P(x)$ can be computed as
\begin{equation}
    \max\limits_{x \in \mathcal{X}} P(x) \le \prod\limits_{i = m k + 1, k \in \mathbb{N},i \le n} \max\limits_{x \in \mathcal{X}} \prod\limits_{1 \le j \le m} P_{i + j}(x) ,
    \label{eq:P_x_upper_bound}
\end{equation}

\noindent where $P_i(x) = Pr(M_i(x) = y_i | x)$. The right hand side of Eq. \eqref{eq:P_x_upper_bound} organizes the queries into groups of size $m$, calculates the $\max P(x)$ within each group, and utilizes the product of these maxima across all groups as the upper bound for the $P(x)$ of all queries. If the complexity of computing $\max_x P(x)$ is $O(c^N)$, where $c$ is a constant and $N$ is the number of queries, then computing the $\max P(x)$ within each group will be $O(c^m)$, and finding the upper bound of $P(x)$ will be $O(c^m \frac{N}{m}) \sim O(N)$.

\begin{algorithm}
    \caption{Branch and Bound for $\log P(x) = f(x) - g(x)$}
    \label{algo:BnB}
  \begin{algorithmic}[1]
    \State $B^x$: box constraint of $x$
    \State $B_0^\mathbf{Y}$: box constraint of $\mathbf{Y}$ (see Section \ref{sec:bnb}).
    \Function{Branch And Bound}{$f, g, B^x, B_0^\mathbf{Y}, \delta = 0.01$}
    \State  $\mathcal{Q}$: a priority queue storing $\{LB, B^\mathbf{Y} \}$ in ascending order
    \State  $UB_g = 0$ // global upper bound of $f(x) - g(x)$
    \State \Call{Bound}{$B_0^\mathbf{Y}$}
    \While{$\text{peek}(\mathcal{Q}).LB \ge UB_g - \delta$} 
        \State  $B^\mathbf{Y}$ = pop$(\mathcal{Q})$
        \State $B_L^\mathbf{Y}$, $B_R^\mathbf{Y}$ = \Call{Branch}{$B^\mathbf{Y}, B_0^\mathbf{Y}$}
        \State\Call{Bound}{$f, g, B^x, B_L^\mathbf{Y}, UB_g, \mathcal{Q}$}
        \State \Call{Bound}{$f, g, B^x, B_R^\mathbf{Y}, UB_g, \mathcal{Q}$}
    \EndWhile
    \State\Return peek$(\mathcal{Q}).LB$
    \EndFunction

    \State
    \Function{Branch}{$B^\mathbf{Y}$, $B_0^\mathbf{Y}$}
        \State $B_{max}$: longest\_edge = None
        \State $l_{max}$: longest\_length = 0
        \State // iterate each edge
        \For{$B^Y, B_0^Y$ in zip($B^\mathbf{Y}, B_0^\mathbf{Y}$)} 
            \State $l$ = $B^Y$.length / $B_0^Y$.length
            \If{$l_{max} < l$}
                \State $l_{max} = l$
                \State $B_{max} = B^Y$
            \EndIf
        \EndFor
        \State $B_L^\mathbf{Y}, B_R^\mathbf{Y}$ = Split $B^\mathbf{Y}$ in half by $B_{max}$
        \State\Return $B_L^\mathbf{Y}$, $B_R^\mathbf{Y}$
    \EndFunction

    \State
    \Function{Bound}{$f, g, B^x, B^\mathbf{Y}, UB_g, \mathcal{Q}$}
        \State $UB$ = $\min_x f(x) - g(x)$ s.t. $B^x, B^\mathbf{Y}$, using any local DC algorithm, e.g. \cite{shen2016dccp}.
        \State $UB_g = \max\{UB_g,  UB\}$ 
        \State Find $H(x)$ as the concave envelope  of $g(x)$
        \State $LB$ = $\min_x f(x) - H(x)$ s.t. $B^x, B^\mathbf{Y}$.
        \State Insert $\{LB,  B^\mathbf{Y}\}$ into $\mathcal{Q}$
    \EndFunction
  \end{algorithmic}
\end{algorithm}

\begin{table*}[t]
\begin{center}
\begin{tabular}{|c | c | c| c | c | c| c |} 
 \hline
 regression & type & $x_1$ & $x_2$ & $x_3$ & $x_4$ & intercept \\ [0.5ex] 
 \hline
 heart disease & Logistic &  -0.059 & -1.456 & -0.0134 & 0 & 6.177\\   
 \hline
 stroke & Logistic & 0.0761 & 0.0952 & 0 & 0.0163 & -7.989\\   
 \hline
 diabetes & Logistic & 0.0491 & 0 & -0.0091 & 0.1039 & -5.07\\  
 \hline
 sleep & Linear & 0.0855 & 0.4617 & -0.07 & 0 & 12.323\\   
 \hline
\end{tabular}
\end{center}
  \caption{The parameters of the health-checkup-data regressions.}
  \label{tab:health_regression}
\end{table*}

\section{Branch-and-bound for Realized Privacy Loss}
\label{sec:bnb}

This section will adapt the branch-and-bound method \cite{benson1991branch} to compute the realized privacy loss given the executed results of the queries introduced in Section \ref{sec:query_to_dc}. Assume a Bayesian Privacy Filter has accepted a sequence of the aforementioned queries, each of which is denoted as $M_i$, and the corresponding output is denoted as $y_i$. The likelihood function of each query can be expressed as the difference of two convex functions: $\log Pr(M_i(x) = y_i) = f_i(x) - g_i(x)$.

As outlined in Section \ref{sec:query_to_dc}, both $f_i(x)$ and $g_i(x)$ share the same form as $F(h(x))$, where $F(y)$ is a convex function, $y$ is a scalar, and $h(x)$ is a linear transformation (i.e., $\theta x + c$). Given any range $[a, b]$ such that $a \le h(x) \le b$, we have

\begin{equation}
    F(h(x)) \le \frac{F(b) - F(a)}{b - a} (h(x) - a) + F(a)  \notag
\end{equation}

\noindent because of the convexity of $F(y)$. The right hand side is denoted as $H(x)$. Since $H(x)$ is a linear transformation of $h(x)$, it is also a linear function of $x$. Notably, as $a \rightarrow b$, $F(h(x)) \rightarrow H(x)$. Consequently, $H(x)$ can act as a concave envelope of $F(h(x))$.

The log-likelihood function $\log P(x)$ of all accepted queries can be represented as $\sum_i f_i(x) - g_i(x)$. Let $g(x) = \sum_i g_i(x)$, and $H_i(x)$ to be the aforementioned concave envelope of $g_i(x)$. A concave envelope of $g(x)$ can then be represented as $\sum_i H_i(x)$. When solving $\min_x \log P(x)$, $f(x) - \sum_i H_i(x)$ remains convex and serves as the lower bound of $\log P(x)$. This process is also applicable to maximizing $\log P(x)$ because it is equivalent to minimizing $g(x) - f(x)$.

With the concave envelope $\sum_i H_i(x)$ in hand, the branch-and-bound method \cite{benson1991branch} can be used to optimize the lower bound of $\log P(x)$. However, searching in the space of $x$ may be slow in practice, prompting an upgrade to the method. Recall that each $f_i(x)$ and $g_i(x)$ can be represented as $F(h(x))$, and if we denote $y = h(x)$, an insightful heuristic emerges — the change of $f_i(x)$ and $g_i(x)$ is more sensitive to $y$ than $x$. Consequently, in the design of the branch-and-bound method, branching (splitting) on the $y$ values of individual queries is favored over branching on $x$.  Let $\mathbf{Y} = \{Y_1, Y_2, ..., Y_n\}$, where $Y_i$ is a variable representing the output of $h_i(x)$. Given that $x$ is bounded, each $Y_i$ possesses its own lower and upper bounds, forming a box constraint denoted as $B_0^{\mathbf{Y}}$, where the constraint of one dimension does not depend on another dimension. The algorithmic details are presented in Algorithm \ref{algo:BnB}.

Notably, the complexity of Algorithm \ref{algo:BnB} increases exponentially with the size of $\mathbf{Y}$. Appendix \ref{sec:group_queries} introduces a strategy to partition the accepted queries into groups of constant size, potentially reducing the overall complexity to linearity.

\section{Realized Privacy Loss of Regressions against Health Checkup Data}
\label{sec:privacy_loss_of_health_data}

A group of individuals possesses health checkup data, including age ($x_1$), sex ($x_2$), blood pressure ($x_3$), and BMI ($x_4$). Table \ref{tab:health_regression} outlines the parameters of the trained regressions.

To ensure privacy, the regressions adhere to LDP guarantee, as outlined in Sections \ref{sec:LDP_of_continuous_value} and \ref{sec:algo_to_compose}. Logistic regressions are discretized to output 0 or 1, and the linear regression for sleep duration is truncated to output 0 or 12. Input data ranges are derived from public datasets: age assumed to range from 10 to 100; sex categorized as 0 (female) or 1 (male); blood pressure varying between 50 and 200; and BMI spanning from 10 to 50. Employing the branch-and-bound method (Section \ref{sec:bnb}), the realized privacy loss of these four regressions is computed for all possible outputs as follows:

\begin{center}
\begin{tabular}{|c | c | c| c | c |} 
 \hline
   $o_1$ & $o_2$ & $o_3$ & $o_4$ & realized privacy loss \\
 \hline
    0 & 0 & 0 & 0 & 2.4639 \\
    \hline
    0 & 0 & 1 & 0 & 2.4084 \\
    \hline
    0 & 0 & 0 & 12 & 1.8036 \\
    \hline
    0 & 0 & 1 & 12 & 2.7253 \\
    \hline
    0 & 1 & 0 & 0 & 2.6865 \\
    \hline
    0 & 1 & 1 & 0 & 3.1550 \\
    \hline
    0 & 1 & 0 & 12 & 2.4642 \\
    \hline
    0 & 1 & 1 & 12 & 3.7449 \\
    \hline
    1 & 0 & 0 & 0 & 3.4761 \\
    \hline
    1 & 0 & 1 & 0 & 2.2610 \\
    \hline
    1 & 0 & 0 & 12 & 2.7511 \\
    \hline
    1 & 0 & 1 & 12 & 2.1975 \\
    \hline
    1 & 1 & 0 & 0 & 2.3362 \\
    \hline
    1 & 1 & 1 & 0 & 1.6863 \\
    \hline
    1 & 1 & 0 & 12 & 1.9062 \\
    \hline
    1 & 1 & 1 & 12 & 2.4959 \\
    \hline
\end{tabular}
\end{center}

\noindent where $o_1, o_2, o_3, o_4$ denote the outputs of the regressions of heart disease, stroke, diabetes, and sleep duration, respectively.

\end{document}